\documentclass[trackchanges,twocolumn]{aastex701}

\usepackage{subcaption}
\usepackage{comment}
\usepackage{amsmath}

\usepackage[left]{lineno}

\begin{document}

\title{FPCA-Enhanced Simulation-Based Inference for Robust Type Ia Supernova Cosmology}

\author[orcid=0000-0001-5197-0363, sname='North America']{Moonzarin Reza}
\affiliation{Department of Physics and Astronomy, Baylor University}
\affiliation{Department of Physics and Astronomy, Texas A\&M University}
\email[show]{moon\_reza@baylor.edu}  
\affiliation{George P. and Cynthia Woods Mitchell Institute for Fundamental Physics and Astronomy, Texas A\&M University}

\author[orcid=0000-0001-7092-9374, sname='North America']{Lifan Wang} 
\affiliation{Department of Physics and Astronomy, Texas A\&M University}
\email{lifan@tamu.edu}  
\affiliation{George P. and Cynthia Woods Mitchell Institute for Fundamental Physics and Astronomy, Texas A\&M University}

\begin{abstract}

Precision cosmology is a cornerstone of modern astronomy, and upcoming wide-field photometric surveys make it crucial to develop robust, data-driven inference methods that can handle complex survey systematics without relying on closed-form likelihoods. Simulation-Based Inference (SBI) meets this need by enabling forward simulations that encode  complex  survey characteristics. Previous SBI analyses in supernova cosmology have used SALT2 light-curve parameters as summary statistics; however, SALT2 is constrained to a two-basis spectral template, imposing rigid modeling assumptions, motivating  a more flexible representation. In this work, we present the first application of Functional Principal Component Analysis (FPCA) light-curve parameters as summary statistics for SBI-based cosmological inference under a non-flat $\Lambda$CDM model. On identically generated simulations, FPCA+SBI yields constraints comparable to both SALT2-based SBI and explicit-likelihood analyses, while providing demonstrably more robust constraints on out-of-domain simulations than SALT2.  Applying the model trained on LSST-like light curves to a spectroscopically confirmed DES Year~5 supernova sample, we recover constraints consistent with those of the DES collaboration to within $0.12\,\sigma$ and $0.23\,\sigma$ in $\Omega_m$ and $\Omega_\Lambda$ respectively, establishing the method's generalizability to real survey data. By introducing host-dependent systematics through a mass-dependent dust extinction law,  we find that FPCA-based summary statistics implicitly encode host-dependent systematics, suggesting that dedicated host modeling may not be necessary within this framework. Combined with its previously demonstrated effectiveness for photometric classification, FPCA offers a compelling foundation for unified, data-driven pipelines capable of jointly performing supernova classification and cosmological inference for upcoming wide-field photometric surveys.

\end{abstract}


\section{Introduction}

The discovery that the deceleration parameter $q$ is negative, established independently by two supernova cosmology teams \citep{riess1998observational, perlmutter1999measurements}, revolutionized cosmology and provided the first direct evidence for the accelerated expansion of the Universe and the existence of dark energy. Since then, Type Ia supernovae (SNe Ia) have been recognized as standardizable candles for measuring cosmic distances. When combined with independent redshift measurements, these distances constrain cosmological 
parameters such as the matter density $\Omega_m$ and the dark energy density $\Omega_\Lambda$. Traditionally, standardization is performed via the Tripp relation \citep{tripp1998two}, correcting the observed peak magnitude using light-curve stretch $x_1$ and color $c$. However, recent studies 
indicate that SNe Ia luminosities retain residual correlations with host galaxy properties --- including stellar mass, star-formation rate, and morphology --- even after conventional light-curve corrections \citep{sullivan2010dependence, rigault2020strong, pruzhinskaya2020dependence}, posing an ongoing challenge for precise cosmological inference.

Traditionally, such analyses are carried out within a Bayesian framework \citep{rubin2025union}, combining a prior with an explicit likelihood via MCMC sampling to obtain the posterior over cosmological parameters. In practice, however, this likelihood is often analytically intractable, forcing 
Gaussian approximations that are inconsistent with the non-Gaussian nature of photometric noise, selection effects, measurement errors Furthermore, upcoming large photometric surveys --- the Vera C.\ Rubin Observatory/LSST 
\citep{ivezic2019lsst} and the Nancy Grace Roman Space Telescope \citep{rose2021reference, akeson2019wide} --- will deliver millions of SNe Ia with heterogeneous cadence, depth, and noise properties, making it increasingly difficult to model diverse survey systematics  in a principled way.

A straightforward likelihood-free approach is approximate Bayesian computation (ABC) \citep{beaumont2019approximate}, which accepts parameter samples only when simulated summary statistics fall within a tolerance $\epsilon$ of the observations --- a scheme that suffers from slow convergence and becomes prohibitively expensive in high dimensions. Simulation-based inference (SBI) \citep{cranmer2020frontier} replaces this rejection scheme with a neural density estimator that learns the mapping between simulator outputs and parameters. Concretely, SBI generates a large set of forward simulations, then trains a conditional density estimator to approximate the posterior  given summary statistics (observed data) --- yielding  a far more sample-efficient and scalable alternative to both classical ABC and explicit likelihood-based methods.

SBI has already been applied to address various systematics in supernova cosmology.  
SIDE-real \citep{karchev2024side} applied truncated marginal neural ratio estimation, a form of SBI, to  SNe Ia light curves, marginalising over thousands of latent variables to infer SN Ia absolute magnitudes and host-galaxy dust properties at the population level. FlowSN \citep{boyd2026flowsn} instead used normalising flows to learn the selection-dependent SN likelihood directly from forward simulations, decoupling it from the distance modulus---and hence the cosmology---so that the learnt likelihood is modular and can be reused across different cosmological models without retraining. CIGaRS \citep{karchev2025cigars} incorporated host-galaxy photometry and physics-based prescriptions for star formation and chemical evolution (Prospector-$\beta$), together with SN~Ia photometry, to infer the intrinsic dependence of SN~Ia brightness on progenitor properties such as age and metallicity; the model also delivers precise and robust host-galaxy photometric redshifts. For direct cosmological parameter inference, existing SBI approaches have used SALT2 \citep{guy2007salt2} light-curve fit parameters --- namely $x_1$, $c$, and $m_b$ --- as summary statistics. \citet{karchev2023sicret} applied Truncated Marginal Neural Ratio Estimation to infer $\Omega_m$ and $\Omega_\Lambda$ in a non-flat $\Lambda$CDM cosmology, while \citet{wang2021likelihood} employed a denoising encoder combined with a Masked Autoregressive Flow for a similar task, further validating their approach on 1048 SNe Ia from the Pantheon compilation \citep{scolnic2018complete}. However, these approaches are fundamentally limited by their reliance on SALT2 parameters, which are derived from a fixed, template-based model with a two-parameter basis --- potentially discarding information encoded in the full light-curve morphology.

This motivates a fully data-driven approach to light-curve compression. Functional Principal Component Analysis (FPCA) directly decomposes observed light curves into a set of data-driven basis functions, capturing subtle morphological variations and complex systematics without imposing a parametric template. FPCA has previously been shown to yield successful photometric classification of SNe Ia \citep{reza2025fpca}; in this work, we present a proof-of-concept analysis demonstrating that FPCA coefficients can serve as summary statistics for SBI-based cosmological parameter inference in a non-flat $\Lambda$CDM cosmology. This also offers a unified framework: since only a single set of light-curve model parameters is required, uncertainties propagate through one model fit rather than two separate pipelines. In contrast, recent photometric cosmology analyses --- DEY5  analysis \citep{abbott2024dark},  LSST forecasting analysis based on ELAsTiCC  \citep{mitra2025fully},  Roman forecasting analysis based on HLTDS simulations \citep{kessler2025cosmology} --- employ separate classification and cosmology steps, using SCONE \citep{qu2021scone} for photometric classification combined with BBC and MCMC for cosmological inference. The FPCA-SBI approach bypasses SALT2 entirely, and naturally lends itself to combining classification and cosmological inference within a single, coherent framework.

We generate cosmological simulations spanning $\Omega_m$ and $\Omega_\Lambda$, sampling all other light-curve parameters consistent with the literature, and apply the Tripp relation to simulate SNe Ia light curves with LSST-like cadence, filters, and noise using \textsc{SNCosmo}. The light curves are fitted with FPCA, and the resulting coefficients are passed through an encoder for dimensionality reduction before being fed into a Mixture Density Network (MDN) that learns the probabilistic mapping between summary statistics and cosmological parameters. We compare the performance of our FPCA-SBI pipeline against both a likelihood-based MCMC baseline and a SALT2-SBI approach, finding consistent posterior biases and uncertainties across all three methods We further compare the robustness of SALT2-SBI and FPCA-SBI on out-of-domian simulations designed to test generalization beyond the training distribution. As a demonstration on real survey data, we apply the trained model to spectroscopically confirmed DES Y5 SNe Ia. Finally, we investigate whether FPCA coefficients inherently capture host galaxy dependencies, or whether explicit host parameter modeling is required for unbiased cosmological inference.

In Section \ref{sec:methods}, we provide a brief overview of simulation-based inference and the functional principal component analysis model. In Section \ref{sec:generation}, we describe the simulation parameters and light curve  generation procedures. In Section \ref{sec:config}, we detail the analysis, including light curve fitting and SBI model training for posterior inference. Finally, in Sections \ref{sec:results} and \ref{sec:conclusions}, we present the results and conclusions, respectively.

\section{Methods Overview}
\label{sec:methods}
\subsection{Simulation-Based Inference}

Simulation-Based Inference consists of 4 building blocks- prior, forward modeling, neural density estimation,   and posterior inference.\\
\textbf{Prior:} This refers to initial knowledge about the (cosmological) parameters we aim to infer. In cosmology with supernova data, this prior can incorporate information from other probes such as the CMB or galaxy clusters, or be chosen to be weakly informative when we wish the data to dominate the constraints.

\textbf{Simulator / Forward Modeling:} With the parameter values  drawn from the prior, the simulator implements the forward model to generate synthetic data vectors or their summary statistics. These summaries can range from a few low-dimensional quantities, such as supernova light curve parameters, to high-dimensional objects such as images.

\textbf{Neural Density Estimation:} A density estimator is trained  on simulated pairs of parameters and corresponding summary statistics to learn the probabilistic mapping,
p($\theta \mid x$), linking light curves to cosmology.

\textbf{Posterior Inference}: The observed data vector (supernova light curves), when fed into the trained estimator, yields posterior samples corresponding to the input cosmology.

The three most common forms of density estimation techniques are Neural Posterior Estimation (NPE), Neural Likelihood Estimation (NLE) and Neural Ratio Estimation (NRE). In NPE, the posterior distribution \(p(\theta \mid x)\) is directly approximated by a neural network. Once trained, the network can be conditioned on new observations to yield constraints on the cosmological parameters. NLE learns an  approximation to the likelihood \(p(x\mid \theta)\) which is then combined with a prior and standard sampling methods such as Markov chain Monte Carlo (MCMC) to obtain samples from the posterior \(p(\theta\mid x)\), without needing an analytic expression for the likelihood. NRE learns the likelihood-to-evidence ratio
\(r(\theta, x) \propto p(x \mid \theta) / p(x)\),
which quantifies how compatible a given parameter--data pair is compared to a random pairing. By multiplying this ratio with the prior and using a sampler such as MCMC, one obtains samples from the posterior \(p(\theta\mid x\)).

Different architectures can be used to model the three density neural estimators described above. Examples include Mixture Density Network (MDN), Masked Autoregressive Flow (MAF), Multi-Layer Perception (MLP), Residual Network (RESNET), Masked Autoencoder for Distribution Estimation (MADE). We only describe  MDN as this is the architecture adopted for the main analysis of this study.  A Mixture Density Network \citep{bishop1994mixture} combines a standard feed-forward neural network with a finite mixture model, typically a mixture of Gaussians.
Given an input vector \(x \in \mathbb{R}^D\), the MDN outputs the parameters of a
mixture of \(K\) Gaussian components: mixture weights
\(\{\pi_k(x)\}_{k=1}^K\), means \(\{\boldsymbol{\mu}_k(x)\}_{k=1}^K\), and
covariance matrices \(\{\boldsymbol{\Sigma}_k(x)\}_{k=1}^K\).
The resulting conditional density is
\begin{equation}
  p( y\mid x)
  \;=\;
  \sum_{k=1}^K \pi_k(x)\,
  \mathcal{N}\!\bigl(y \mid \boldsymbol{\mu}_k(x), \boldsymbol{\Sigma}_k(x)\bigr),
\end{equation}
where \(\mathcal{N}(y \mid \boldsymbol{\mu}, \boldsymbol{\Sigma})\) denotes a (typically multivariate)
Gaussian density.

The MDN is trained by maximizing the  log-likelihood of the observed targets under this mixture model. In the context of SBI, MDNs are frequently used as flexible
parametric models for conditional densities such as \(p(\theta \mid x)\) (for NPE),
\(p(x \mid \theta)\) (for NLE), or related quantities, by choosing \(y\) to be the
quantity of interest and \(x\) the conditioning variable.

\subsection{Functional Principal Component Analysis}
\label{sec:fpca_first}
Functional Principal Component Analysis (FPCA) is a light-curve fitting technique designed to model  sparse or irregularly-sampled light curves \citep{he2018characterization}. Each light curve is represented as the weighted summation of basis functions/ principal components. These FPCA basis functions are derived using well-sampled light curves from Harvard- Smithsonian Center for Astrophysics \citep{hicken2009cfa3, hicken2012cfa4}, Lick Observatory Supernova Search \citep{ganeshalingam2010results}, and Carnegie Supernova Project \citep{contreras2010carnegie}. The coefficients of principal components are called scores and are used to establish relations between light curve shapes and physical quantities such as intrinsic color, spectral line strength, etc. The basis functions are data-driven mathematical constructs,  and as a result can capture more subtle variations in the temporal evolution of light curves beyond what rigid template-fitting models such as SALT2 \citep{guy2007salt2} can capture. 

\begin{figure*}
    \centering
    \includegraphics[width=\linewidth]{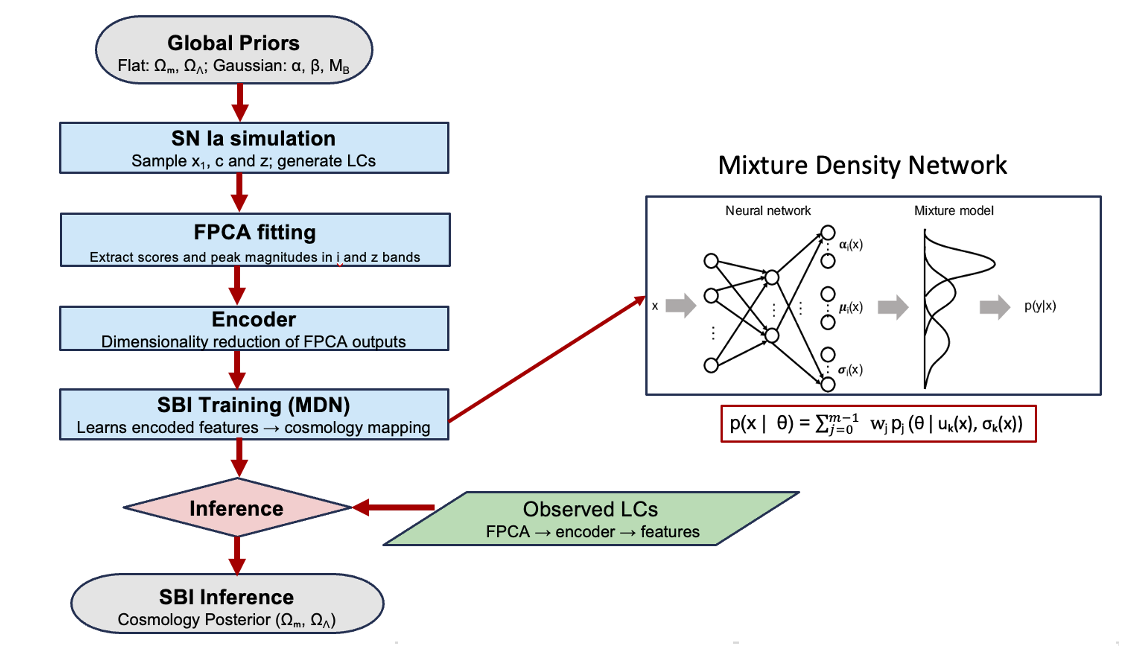}
    \caption{Flowchart of the cosmological analysis pipeline implemented in this work. Global priors (flat for cosmology, Gaussian for light‑curve nuisance parameters), combined with SALT2 model is used to simulate SN Ia light curves, which are then fitted using FPCA. The resulting FPCA parameters are passed into an enocder that compresses them into a lower-dimensional latent space. The encoded features, together with the cosmological parameters, are then used to train a neural network that learns the mapping between the latent features and cosmology.}
    \label{fig:block_diag_sbi_this_work}
\end{figure*}

\section{Data generation}
\label{sec:generation}
\subsection{Overall Pipeline}
In Figure \ref{fig:block_diag_sbi_this_work}, we show the flowchart of the SBI pipeline adopted for the cosmological analysis in this study. The global priors consist of cosmological parameters \((\Omega_m, \Omega_\Lambda)\) and nuisance parameters \((\alpha, \beta, M_B)\). We generate 2000 sets of global cosmology  simulations for the final training set. For the test set, we simulate 100 sets of simulations. For each set of cosmology, 6000 light curves are simulated by sampling the light curve parameters \((x_1, c, z)\). The light curves are then fitted using the FPCA framework described in Section \ref{sec:fpca_first}. To reduce the dimensionality of the feature space, we use an encoder to project the FPCA model-fit parameters into a lower-dimensional latent space. A mixture density network then learns the probabilistic mapping between the latent (encoded) features and the underlying cosmological parameters. Once trained, the network infers the posterior distribution over cosmological parameters for  new sets of observed light curves.
\subsection{Parameters}
The Tripp relation \citep{tripp1998two} is used to standardize the absolute luminosity of Type Ia supernovae, and forms the foundation of  supernova cosmology.  It is given by: 

\begin{equation}
m_B = M_B  - \alpha x_1 + \beta c + \mu(z; \Omega_m, \Omega_\Lambda)
\label{eq:tripp}
\end{equation}
 where $m_B$ denotes the apparent peak magnitude in the \(B\) band,   $M_B$ is the  absolute $B$-band magnitude,  $x_1$ is the light-curve stretch parameter, $c$ is the color parameter, $\alpha$ is the stretch-luminosity slope, The coefficients $\alpha$ and $\beta$
 are the stretch–luminosity and color–luminosity slopes, respectively, and the term \(\mu (z; \Omega_m, \Omega_\Lambda)\) is the distance modulus for a given cosmology and encodes the cosmological information. Assuming a $\Lambda$CDM cosmological model, we generate a suite of simulations in which the quantities on the left-hand side of Equation~\ref{eq:tripp} are drawn from the distributions summarized in Table \ref{tab:priors_training_testing}, which  specifies the cosmological, nuisance, and light-curve fit parameters.  The apparent magnitude $m_B$ for each simulated supernova is then computed using the Tripp relation. 
 
 As described in Table \ref{tab:priors_training_testing} the cosmological parameters, $\Omega_m$ and $\Omega_\Lambda$, are sampled from broad uniform priors  for the training set. For the test set, these parameters are instead drawn from  Gaussian priors centered around the Planck values $(0.3,\ 0.7)$  \citep{aghanim2020planck}, in order to evaluate the model's performance on 
cosmologically realistic parameter values.

The priors for the global nuisance parameters and the light curve fit parameters  are motivated by existing literature on SN Ia cosmology.  Truncated normal distributions are adopted  for $\alpha$, $\beta$, and $M_B$, with identical means and standard deviations across both sets; however, the test simulations  employ narrower truncation bounds in order to focus on more realistic ranges \citep{2018ApJ...859..101S}. Similarly, the 
light curve fit parameters are sampled from truncated normal distributions, with  identical distributions applied to both the training and test sets following the treatment in Pantheon+ \citep{scolnic2022pantheon+}.

\begin{table*}
\centering
\begin{tabular}{c|c|c}
\hline
Parameter & Prior (Training) & Prior (Test)\\
\hline
$\Omega_m$ & $\mathcal{U}(0,1)$ & $\mathcal{TN}(0.3,\,0.1;\,0.0,\,0.6)$\\
\hline
$\Omega_\Lambda$ & $\mathcal{U}(0,1)$  &
$\mathcal{TN}(0.7,\,0.1;\,0.4,\,1.0)$ \\
\hline
$\alpha$ & $\mathcal{TN}(0.167,\,0.012;\,0.131,\,0.203)$ 
& $\mathcal{TN}(0.167,\,0.012;\,0.143,\,0.191)$\\
\hline
$\beta$ & $\mathcal{TN}(3.51,\,0.16;\,3.03,\,3.99)$
&  $\mathcal{TN}(3.51,\,0.16;\,3.19,\,3.83)$ \\
\hline
${M_B}$ 
& $\mathcal{TN}(\mu_{M_B},\,\sigma_{M_B};\,-19.8,\,-18.8)$ & $\mathcal{TN}(\mu_{M_B},\,\sigma_{M_B};\,-19.6,\,-19.0)$ \\
& \quad $\mu_{M_B} \sim \mathcal{N}(-19.3,\,0.1)$  & \quad $\mu_{M_B} \sim \mathcal{N}(-19.3,\,0.1)$\\
& \quad $\sigma_{M_B} \sim \mathcal{U}(0.08,\,0.12)$ & \quad $\sigma_{M_B} \sim \mathcal{U}(0.08,\,0.12)$\\
\hline
$x_1$ & $\mathcal{TN}(0,\,1;\,-3,\,3)$
&$\mathcal{TN}(0,\,1;\,-3,\,3)$\\
\hline
$c$ & $\mathcal{TN}(0,\,0.1;\,-0.3,\,0.3)$
& $\mathcal{TN}(0,\,0.1;\,-0.3,\,0.3)$\\
\hline
\end{tabular}
\caption{Prior distributions of cosmological, nuisance and light curve  parameters for the training set and test set used for the simulations. 
Here $\mathcal{U}$, $\mathcal{N}$, and $\mathcal{TN}$ denote Uniform, Normal, 
and Truncated Normal distributions, respectively.}
\label{tab:priors_training_testing}
\end{table*}

\subsection{Redshift Sampling from Comoving Volume and Supernova Rate}

\subsubsection{Cosmology and distance measures}
We follow the formalism of \citep{hogg1999distance} to compute the comoving volume in redshift bins. Assuming a Friedmann--Lemaître--Robertson--Walker cosmology with matter density $\Omega_{\mathrm{m}}$, cosmological constant $\Omega_{\Lambda}$, curvature $\Omega_{k} = 1 - \Omega_{\mathrm{m}} - \Omega_{\Lambda}$, and Hubble constant $H_{0}$, we evaluate the comoving volume element as a function of redshift.

The dimensionless Hubble parameter is given by:
\begin{equation}
E(z) = \frac{H(z)}{H_0} = \sqrt{\Omega_{\mathrm{m}}(1+z)^{3} + \Omega_{k}(1+z)^{2} + \Omega_{\Lambda}}
\end{equation}

The line-of-sight comoving distance is given by:
\begin{equation}
\chi(z) = \frac{c}{H_{0}} \int_{0}^{z} \frac{dz'}{E(z')}
\end{equation}
The transverse comoving distance is given by:

\begin{equation}
D_{\mathrm{M}}(z) =
\left\{
\small
\begin{array}{ll}
\chi(z), & \Omega_{k} = 0,\\[8pt]
\dfrac{c}{H_{0}\sqrt{\Omega_{k}}}\,\sinh\!\Big(\sqrt{\Omega_{k}}\,H_{0}\chi(z)/c\Big), & \Omega_{k} > 0,\\[10pt]
\dfrac{c}{H_{0}\sqrt{|\Omega_{k}|}}\,\sin\!\Big(\sqrt{|\Omega_{k}|}\,H_{0}\chi(z)/c\Big), & \Omega_{k} < 0.
\end{array}
\right.
\end{equation}

\subsubsection{Comoving volume element}

The comoving volume element per unit redshift and solid angle is:
\begin{equation}
\frac{dV_{\mathrm{c}}}{dz\,d\Omega}
= \frac{c}{H_{0}}\,\frac{D_{\mathrm{M}}^{2}(z)}{E(z)}
\end{equation}

\subsubsection{Event rate and redshift distribution}

Let $R(z)$ be the comoving event rate density (per unit comoving volume per unit source-frame time). For the LSST SNe~Ia sample, this quantity is assumed to scale as $(1+z)^{1.5}$ $\mathrm{vol}^{-3}\,\mathrm{t}^{-1}$ \citep{abell2009lsst}. The  observed number of supernovae per unit solid angle
and redshift is\,
\begin{equation}
\frac{d{N}}{dz\,d\Omega}
= \frac{R(z)}{1+z}\,\frac{dV_{\mathrm{c}}}{dz\,d\Omega}
= \frac{R(z)}{1+z}\,\frac{c}{H_{0}}\,\frac{D_{\mathrm{M}}^{2}(z)}{E(z)}
\end{equation}
where the factor \(1/(1+z)\) accounts for cosmological time dilation, since source-frame time intervals are stretched by a factor of \((1+z\)) in the observer frame.
More generally, if we absorb all redshift  factors into
a weight $W(z)$, then the redshift distribution is:
\begin{equation}
\frac{dN}{dz} \propto (1+z)^{\frac{1}{2}}\,\frac{D_{\mathrm{M}}^{2}(z)}{E(z)}
\label{eq:redshift_final_eq}
\end{equation}

The corresponding  redshift probability density  in a redshift bin $[z_1, z_2]$ is:
\begin{equation}
p(z) = \int_{z_1}^{z_{2}}
(1+z')^{1/2}\,D_{\mathrm{M}}^{2}(z')/E(z')\,dz'
\end{equation}

The sky-blue histogram in Figure \ref{fig:simulated_z_dist} shows the redshift distribution of the simulated supernova sample obtained using Equation \ref{eq:redshift_final_eq}, which predicts an increasing number of supernova explosions at higher redshift. This increase is a consequence of the growth of comoving volume with redshift, combined with the assumed volumetric supernova rate, which together predict more explosions per redshift bin at higher redshift.

\subsection{Light Curve Generation, Errors and  Detection}
\subsubsection{Light-curve synthesis}

We use the \texttt{SALT2‑extended model} implemented in the \texttt{sncosmo} Python package \citep{2016ascl.soft11017B} to generate  synthetic Type Ia supernova light curves in  the Rubin Observatory LSST $i$ and $z$ bands, expressed in the AB magnitude system. Taken together, the resulting peak magnitude $m_B$ (Equation \ref{eq:tripp}) and the parameters $(x_1, c, z)$  fully specify the \texttt{SALT2-extended} spectral energy distribution for each simulated supernova.  Observer frame fluxes (corresponding to rest-frame phases from $-15$ to $+45$ days relative to $B$-band maximum light), are then simulated with a fixed  4-days cadence in each band.  This fixed cadence is a simplifying assumption, and in future work we plan to incorporate more realistic survey cadences and observing conditions.

\subsubsection{Noise and photometric uncertainties}
We model photometric uncertainties of the observed magnitudes as described in the LSST Science Book, Chapter~3 \citep{LSSTScienceBook}. The total single-visit photometric uncertainty is:
\begin{equation}
\sigma_1^2 = \sigma_{\rm sys}^2 + \sigma_{\rm rand}^2 
\label{eq:sigma_total}
\end{equation}
where \(\sigma_{\rm sys}\) is the systematic term (LSST calibration system and procedures maintain \(\sigma_{\rm sys}<0.005\)) and \(\sigma_{\rm rand}\) is the random photometric error (limited by sky noise, readout noise, etc) per visit. The random component as a function of magnitude is
\begin{equation}
\sigma_{\rm rand}^2 = (0.04 - \gamma)\,x + \gamma\,x^2 \quad {\rm (mag^2)}
\label{eq:sigma_rand}
\end{equation}
with
\begin{equation}
x = 10^{0.4\,(m - m_5)} .
\label{eq:x_def}
\end{equation}
Here \(m\) is the source magnitude in a given LSST band, \(m_5\) is the single-visit \(5\sigma\) depth in that band, and \(\gamma\) is a band-dependent parameter  listed in Table~\ref{tab:lsst_m5_gamma}.

\begin{table}[t]
\centering
\caption{Single-visit $5\sigma$ depths ($m_5$) and $\gamma$ parameters for LSST bands.}
\label{tab:lsst_m5_gamma}
\begin{tabular}{lcccccc}
\hline
Band          & $g$   & $r$   & $i$   & $z$   & $y$   \\
\hline
$m_5$ (mag)  & 25.0  & 24.7  & 24.0  & 23.3  & 22.1 \\
$\gamma$    & 0.038 & 0.039 & 0.039 & 0.040 & 0.040 \\
\hline
\end{tabular}
\end{table}

\subsubsection{Detection criteria}
We retain only 3‑sigma detections, defined here as measurements with photometric uncertainty $\sigma_1<0.36$, by applying a hard signal-to-noise cut. In future work, we plan to replace this thresholding with a probabilistic detection scheme, in which each observation is assigned a detection probability based on its signal-to-noise ratio and incorporated into the analysis rather than a strict threshold cut.

\begin{figure*}[htbp]
\centering
\begin{subfigure}[t]{0.48\textwidth}
    \centering
    \includegraphics[width=\textwidth]{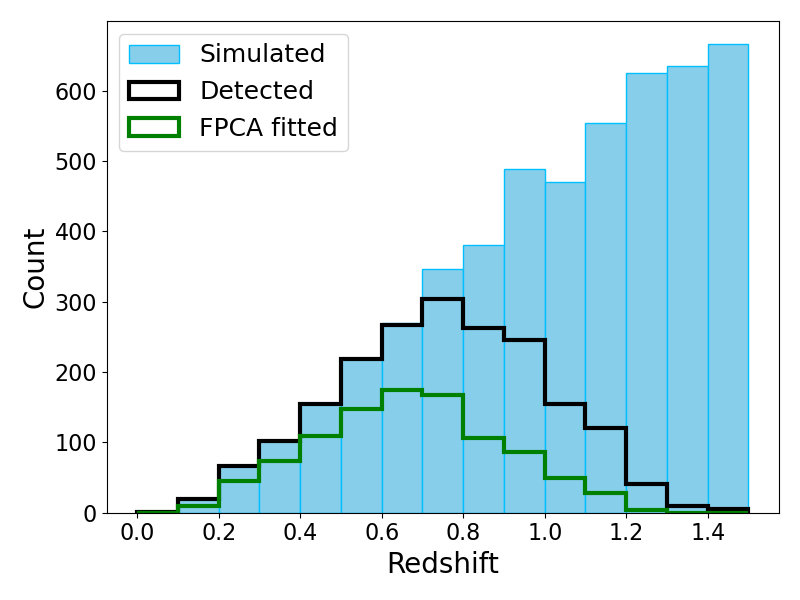}
    \caption{Redshift distribution of the simulated, detected, and FPCA-fitted 
    supernova samples. The blue-filled histogram represents the full simulated 
    population, the black step histogram shows supernovae satisfying the $3\sigma$ 
    detection threshold, and the green step histogram corresponds to supernovae 
    for which the FPCA fit converges within reasonable errors.}
    \label{fig:simulated_z_dist}
\end{subfigure}
\hfill
\begin{subfigure}[t]{0.48\textwidth}
    \centering
    \includegraphics[width=\textwidth]{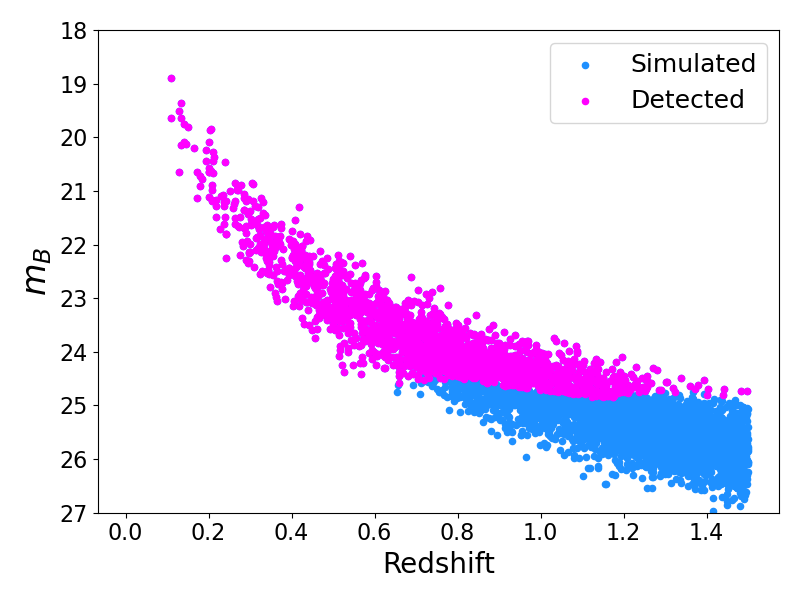}
    \caption{Peak apparent magnitude $m_B$ as a function of redshift for the simulated (blue) and detected (magenta) supernovae sample, where detection is defined 
    by a $3\sigma$ significance threshold.}
    \label{fig:mag_sim_det}
\end{subfigure}
\caption[Redshift distribution and peak apparent magnitude $m_B$ for the simulated and detected supernova samples]{Redshift distribution and peak apparent magnitude $m_B$ for the simulated and detected supernova samples, illustrating the magnitude-limited selection function  and the onset of Malmquist bias beyond 
$z \approx 0.7$.}
\label{fig:redshift_mag_dis}
\end{figure*}

In Figure \ref{fig:simulated_z_dist}, we plot the redshift distribution of the detected supernovae sample as black step histograms overlaid on the simulated supernova blue histograms.
The detection efficiency is high at low redshifts, with nearly all simulated supernovae recovered up to $z \approx 0.6$. Beyond this redshift, the detection fraction starts declining and becomes negligible by $z \approx 1.3$. This falloff is consistent with a magnitude-limited survey, wherein supernovae at greater distances become too faint to satisfy the signal-to-noise threshold. The peak of the detected distribution occurs in the redshift bin $0.7 \leq z \leq 0.8$, after which Malmquist-bias progressively suppresses the detected fraction.

Figure \ref{fig:mag_sim_det} shows the peak apparent magnitude $m_B$ as a function of redshift for both the simulated (blue) and detected (magenta) samples, where detection is defined by a $3\sigma$ significance threshold. The simulated population spans a broad range of apparent magnitudes at each redshift, reflecting variation in cosmological parameters ($\Omega_m, \Omega_\Lambda$) across simulations — supernovae at the same redshift exhibit different apparent magnitudes depending on the underlying cosmology, since the luminosity distance $d_L(z)$
 is sensitive to the cosmological model. Up to $z \approx 0.7$, nearly the entire simulated population is detected, indicating that the survey is effectively complete in this redshift range. Beyond $z \approx 0.7$, the detected sample becomes increasingly confined to the brighter end of the magnitude distribution, illustrating the onset of magnitude-limited selection. A negligible number of supernovae are detected fainter than $m_B \approx 24.5-25$, consistent with the survey's limiting magnitude. This selection effect introduces a Malmquist bias at high redshifts, with the detected sample becoming increasingly incomplete beyond $z>0.8$.

\section{Light Curve Fitting and Model Training}
\label{sec:config}

\subsection{FPCA-based fitting}

Although the principal component basis functions are originally designed to establish connections with intrinsic supernova properties, they have also proven effective for photometric classification \citep{reza2025fpca}. Here, we apply the same FPCA features to cosmological parameter inference, with the goal of developing a unified framework that proceeds directly from light curves to cosmology using a single fitter.

The FPCA model with two principal components for the light curve in the $j$-th band at observation time $t_{jk}$ is given in Equation \ref{eqn:1}:

  \begin{equation}
\begin{aligned}
f_{model}(t_{jk}, z; \Theta_j) &= \hat{m}_j + \phi_{0}(\frac{t_{jk} - \hat{t}_j}{1 + z})  
\\
&+ a_{1j} \phi_{1}(\frac{t_{jk} - \hat{t}_j}{1 + z}) + a_{2j} \phi_{2}(\frac{t_{jk} - \hat{t}_j}{1 + z}), 
\\
\Theta_i &= \{ \hat{m}_j, \hat{t}_j, a_{1j}, a_{2j} \},
\end{aligned}
\label{eqn:1}
\end{equation}

 Here, $\hat{m}_j$ denotes the peak magnitude in the $j$-th band, $\hat{t}_j$ is the corresponding peak time, and $(a_{1j}, a_{2j})$ are the first two FPCA coefficients for that band.  The quantity $t_{jk}$ represents the time of the $k$-th observation in band $j$, and $z$ is the known  redshift. The phase $\frac{t_{jk}-\hat{t}_j}{1+z}$ is defined over the interval $-10$ to $40$ days to restrict the FPCA model to the region where it is trained and to avoid extrapolation. The functions $\phi_0$, $\phi_1$, and $\phi_2$ correspond to the zero-, first-, and second-order principal components of the FPCA model, respectively. The model is fully specified by the parameter vector $\Theta_i$ for each band. We employ a filter-vague FPCA model, in which the principal components are identical across all filters.

We determine the parameter vector $\Theta_j$ for each light curve following Step 1 (Equation~5) of \citet{reza2025fpca}. The model parameters are determined by minimizing residuals between the observed data and model predictions, subject to penalty terms on the two FPCA coefficients, which are calibrated using a clean sample of PLAsTiCC SNe Ia (see \citealt{reza2025fpca} for details). Parameter optimization is performed using the \textsc{mpfit} module \citep{markwardt2009non}, which implements least-squares minimization via the Levenberg--Marquardt algorithm.

For this analysis, we fit only those light curves with at least 5 3-$\sigma$ observations in each of $i$ and $z$ bands. Furthermore, we retain only those light curves with redshift $z > 0.05$, to exclude peculiar velocity effects and focus on the Hubble flow regime, and for which the uncertainties in the estimated parameters and the goodness of  fits are within acceptable bounds. From this subset, the first 800 light curves satisfying the above criteria are retained per cosmology. The choice of selecting 800 SNe Ia per simulated realization is motivated by the fact that it is broadly comparable to the scale of current-generation SN Ia cosmology samples (Dark Energy Survey). This choice is a conservative one, balancing computational cost against providing a stable proof-of-concept test of the pipeline.

Figure~\ref{fig:simulated_z_dist} shows the redshift distribution of the FPCA-fitted light curves as green step histograms. At $z > 0.8$, an increasingly smaller fraction of light curves satisfy the quality cuts.

\begin{figure*}[htbp]
    \centering

    \begin{subfigure}{0.48\textwidth}
        \centering
        \includegraphics[width=\linewidth]{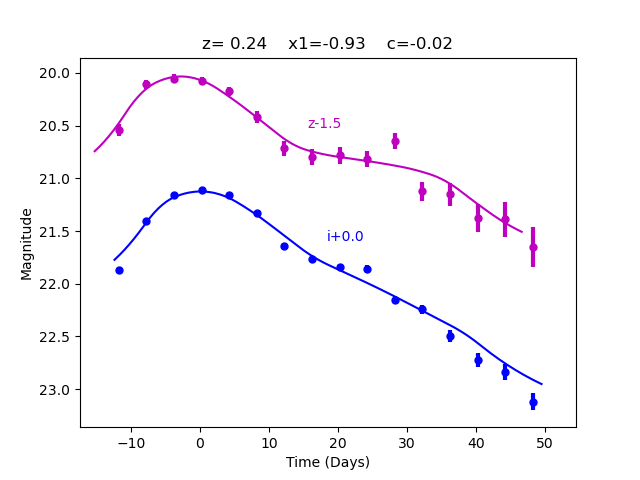}
    \end{subfigure}
    \hfill
    \begin{subfigure}{0.48\textwidth}
        \centering
        \includegraphics[width=\linewidth]{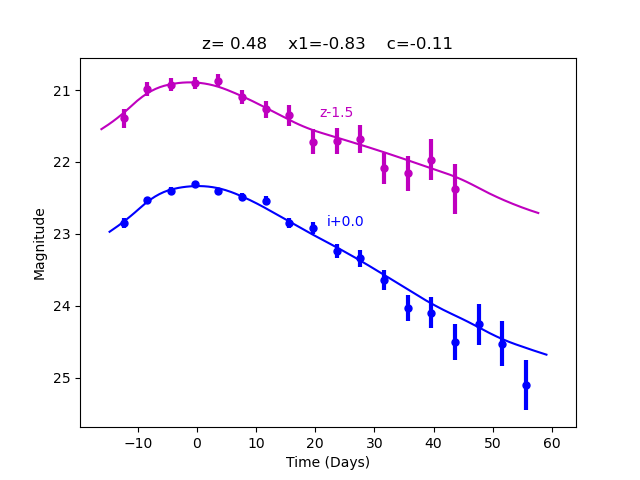}
    \end{subfigure}
    
    \vspace{0.3cm}
    
    \begin{subfigure}{0.48\textwidth}
        \centering
        \includegraphics[width=\linewidth]{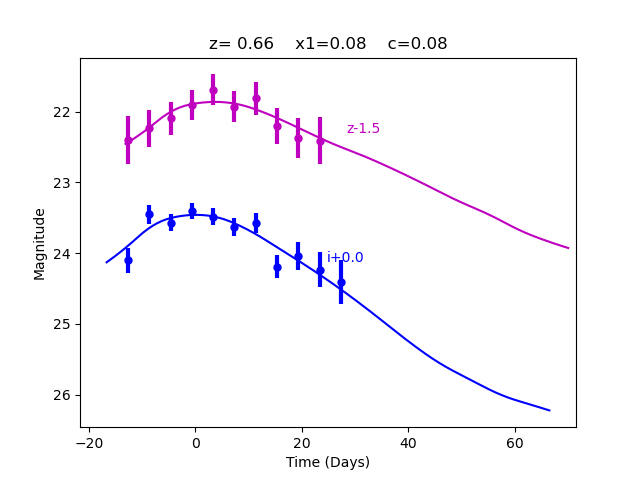}
    \end{subfigure}
    \hfill
    \begin{subfigure}{0.48\textwidth}
        \centering
        \includegraphics[width=\linewidth]{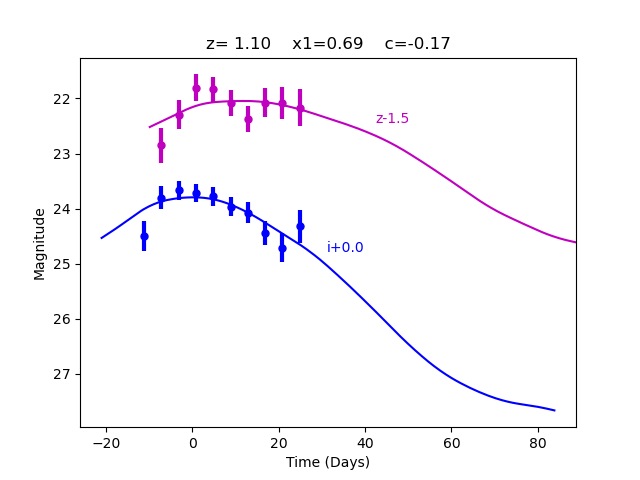}
    \end{subfigure}
  
    \caption{Four example FPCA light curve fits at redshifts $z = 0.24$, $z=0.48$,
$z = 0.66$, and $z=1.10$. The 
$i$-band and $z$-band fits are shown in blue and purple respectively, with the 
$z$-band offset by $-1.5$.  magnitudes for clarity. Data points with error bars are shown alongside the FPCA model fit. The redshift $z$, stretch parameter $x_1$,
 and color parameter $c$ for each supernova are indicated in the subplot titles.}
    \label{fig:lc_fits_4}
\end{figure*}

Figure~\ref{fig:lc_fits_4} shows representative fits for four light curves at redshifts $z = 0.24$,
$z = 0.48$, $z= 0.66$, and $z = 1.10$.   The redshift $z$, stretch parameter $x_1$, and color parameter $c$ are shown as the title of each subplot. The $i$-band fit is shown in blue and the $z$-band fit in purple.

For each cosmology, we ultimately retain 800  SNe that pass our analysis cuts. These choices for the number of simulations (2000) and SNe per simulation are motivated by a convergence study based on 100 dedicated test simulations, in which we use the simulated (true) SALT2 parameter values  during model training. In this framework, we monitor how the 1-$\sigma$ uncertainty on the cosmological parameters fluctuates as we increase both the number of simulations and the number of SNe, and we find that with 2000 simulations and 800 light curves per simulation the uncertainty has converged to a stable plateau.

We now describe the structure of the encoder which is used to compress the FPCA parameters for dimensionality reduction.

\subsection{Dimensionality Reduction using Encoder}
\label{sec:encoder}
\begin{figure}[htbp]
    \centering
    \includegraphics[scale=0.42]{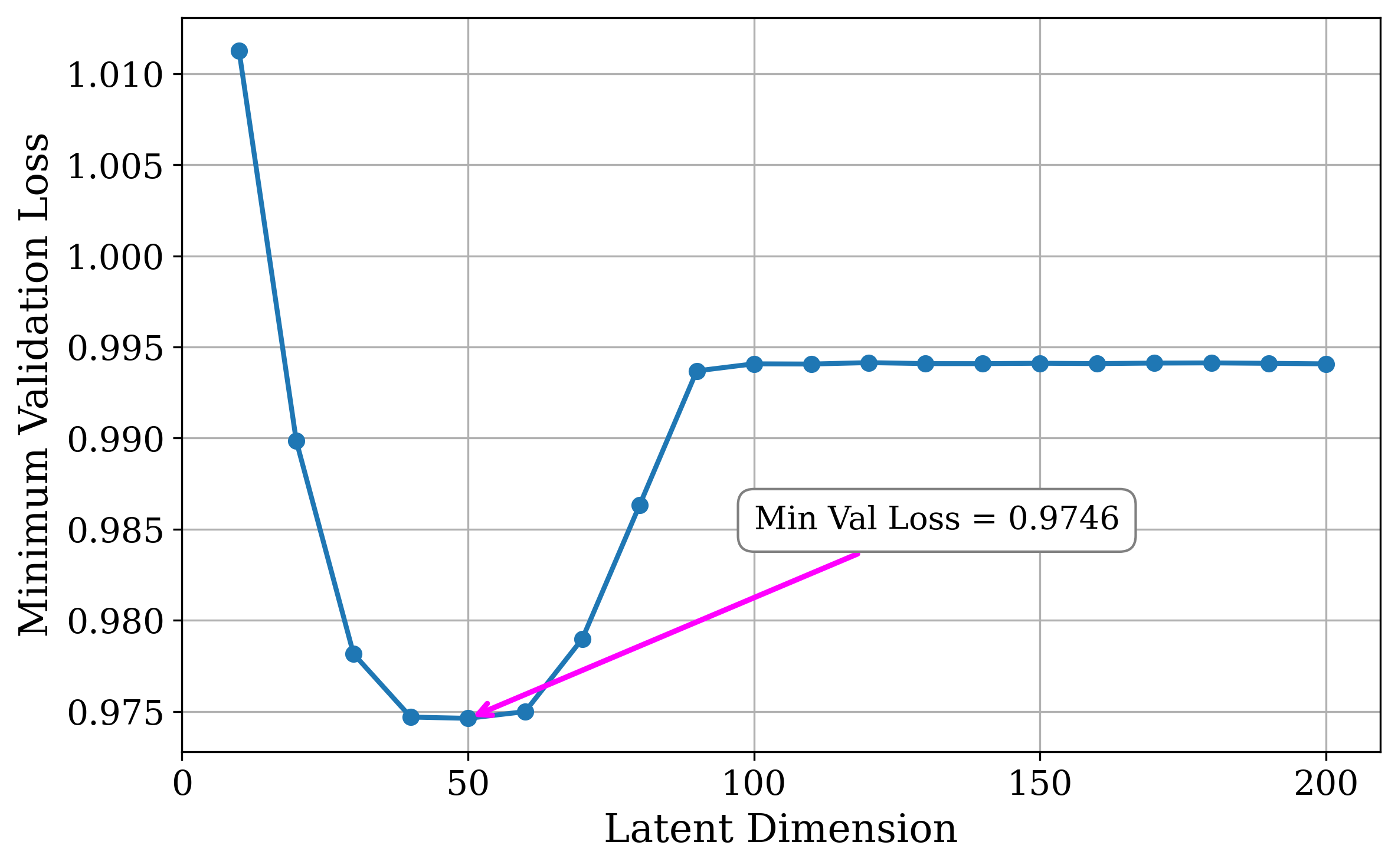}
    \caption[Encoder validation loss versus latent dimension]{Minimum validation loss (MSE) as a function of latent space dimensionality for a feedforward encoder with two fully connected layers  and ReLU activation, trained via autoencoder reconstruction loss. The loss is minimised (0.9746) at a latent
dimension of 50, which is adopted for all subsequent analysis.}
    \label{fig:loss_dim}
\end{figure}

\begin{figure}[h]
    \centering
    \includegraphics[scale=0.52]{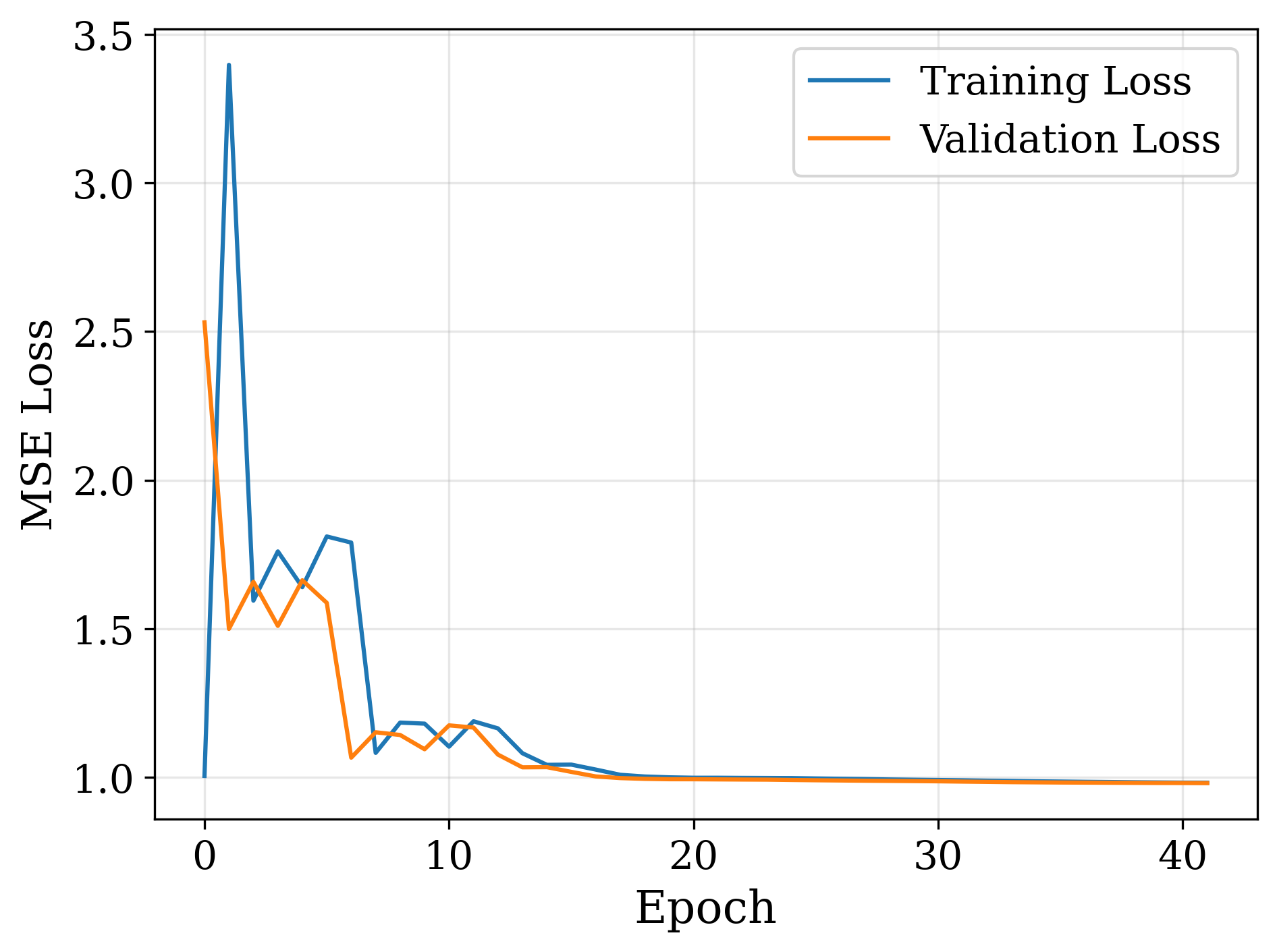}
    \caption[Encoder training and validation loss]{Training and validation MSE loss
as a function of epoch for the encoder with a latent dimension of 50. Both losses converge within 20 epochs with no sign of overfitting.}
    \label{fig:enc_loss}
\end{figure}

Encoders are widely used in high-dimensional machine learning problems to reduce the dimensionality of the feature space, making training faster and more stable, while also removing redundant or correlated features, mitigating overfitting, and revealing latent structure in the data.

We use an autoencoder to compress the FPCA summary statistics prior to inference. In its original form, the feature vector for each supernova consists of the peak apparent magnitude and two FPCA scores in each of the $i$- and $z$-bands, together with the redshift, yielding 7 features per supernova. For a simulation of 800 SNe Ia, this gives a total input dimensionality of 5,600 features. 

The encoder consists of a feedforward neural network with two fully connected layers. The first layer maps the 5,600-dimensional input to a hidden layer of 1,024 units with a ReLU activation function. The second layer projects the hidden representation to the latent space of dimension $d_{\mathrm{latent}}$. During training, a symmetric decoder mirrors this architecture, mapping the latent representation back through a 1,024-unit hidden layer to reconstruct the original input. The reconstruction loss (MSE) serves as a proxy training objective, ensuring the latent space captures the dominant structure of the input feature vector.  Only the encoder is retained at inference; the decoder is discarded after training. The encoder is trained on 800 cosmology realisations and validated on a held-out set of 200 realisations, used for latent dimensionality selection via minimum validation MSE. The encoder weights are then fixed, and all subsequent analysis uses the fixed latent representations produced by this encoder.  We plot the minimum validation loss (MSE) as a function of latent dimension  in Figure~\ref{fig:loss_dim}. The minimum validation loss is achieved at a latent dimensionality of 50, and we therefore retain this configuration for all subsequent analysis.

Figure~\ref{fig:enc_loss} shows the training and validation MSE loss as a function of epoch for the encoder with a latent dimension of 50. Both losses
converge rapidly within 20 epochs and remain closely tracked throughout training, with no sign of overfitting, indicating that the encoder generalizes well to unseen data. This is notably achieved with only 800 training realisations across a 5,600-dimensional input space, demonstrating that the bottleneck architecture alone provides sufficient regularization without additional techniques such as dropout.

We perform a similar analysis using the SALT feature vector, which consists of four parameters ($m_B$, $x_1$, $c$, $z$) per supernova, yielding a total input dimensionality of $4 \times 800 = 3200$. We retain the same two-layer encoder architecture (1024 units per layer) and find that the reconstruction loss is minimized at a latent dimension of 60, with 0.1\% difference between  50 and 60 dimensions. For consistency with the FPCA-based encoder, we adopt a 50-dimensional 
latent space for SALT-based summary statistics as well.

\subsection{SBI Configuration}

In this work, we use Sequential Neural Posterior Estimation (SNPE) with a Mixture Density Network (MDN) as the density estimator, using the default configuration of the \texttt{sbi} package~\citep{cranmer2020frontier}. SNPE directly learns the posterior distribution $p(\theta \mid \mathbf{x})$ by
training a conditional density estimator on simulated parameter-data pairs, refining the proposal distribution sequentially across multiple rounds using posterior samples from previous rounds. Once trained, the density estimator can be evaluated at negligible computational cost for any new observation, without retraining. We perform baseline comparisons of computational runtime between different density estimation techniques (NPE, NRE, and  NLE), as well as between different architectures (MDN, MAF, MLP, etc) used to implement the density estimators. MDN and MAF based density estimators in Appendix \ref{app:time_sbi_meth}

\subsection{Computational Resources and Runtime}
All analyses are performed on an Apple M2 chip with 8GB of RAM. The computational cost is dominated by light curve generation and fitting. We generate 6,000 light curves per simulation across the $i$- and $z$-bands, giving 12,000,000 light curves in total for 2,000 training simulations, which takes approximately 3 hours. Fitting the FPCA basis to both bands across 800 light curves per simulation yields 1,600,000 fits (per light curve); and this process requires approximately 22 hours. By comparison, fitting the same
light curves with the SALT2 fitter in \texttt{sncosmo} takes approximately 64 hours, representing an increase in compute time by a factor of almost three. This demonstrates a significant computational advantage of FPCA over SALT2 for large-scale light curve modeling. Training the MDN and  posterior inference on a test simulation is much faster and completes in seconds.

\section{Results}
\label{sec:results}

The standard approach for estimating cosmological parameters from Type~Ia supernovae is to fit the light curves with the SALT2/SALT3 model to obtain the parameters \(m_b\), \(x_1\), and \(c\). These parameters are then related to cosmology through the Tripp relation (Equation~\ref{eq:tripp}), and the cosmological parameters are inferred using explicit likelihood-based sampling with Markov Chain Monte Carlo (MCMC); we refer to this baseline as the \textbf{SALT-MCMC} method.

A simulation-based alternative, which we term \textbf{SALT-SBI}, reflects a recent trend in the field towards likelihood-free inference: forward-simulated supernovae are generated using the Tripp relation, those light curves are fitted with SALT2/SALT3 to obtain \(\{m_b, x_1, c\}\), and a mixture density network (MDN) is trained to learn the mapping from these parameters to the cosmological parameters, bypassing explicit likelihood evaluation.

Our approach, \textbf{FPCA-SBI}, uses the same Tripp-relation-based forward simulations as SALT-SBI but differs in the light-curve fitting step: functional principal component analysis (FPCA) is used instead of SALT2/SALT3, and the resulting FPCA scores are used as summary statistics. These FPCA coefficients are then passed to an MDN to perform likelihood-free posterior inference.

First, we compare SALT-MCMC, SALT-SBI, and FPCA-SBI using raw summary statistics, without an autoencoder-based compression step. Second, we compare SALT-SBI and FPCA-SBI with an autoencoder architecture, assessing whether learned compression improves posterior accuracy and uncertainty calibration. Third, we test generalization by evaluating both SALT-SBI and FPCA-SBI on out-of-domain simulations. Finally, we conduct two analyses using FPCA-SBI only. We apply an MDN trained on simulated LSST-like light curves to the DES~Y5 spectroscopic sample and compare the resulting cosmological constraints with those obtained by the DES team. We also investigate whether FPCA scores can passively capture host-dependent systematics. To this end, we modify the Tripp relation by introducing a host-dependent term and examine whether explicitly including host-galaxy parameters as additional inputs to the MDN yields improved bias and reduced uncertainties in the recovered cosmology.

We assess cosmological parameter recovery using two main metrics: the median posterior uncertainty, \(\sigma_{\theta}\), and the median posterior bias, \((\hat{\theta} - \theta_{\mathrm{true}})/\sigma_{\theta}\), where $\hat{\theta}$ is the median of the marginalized posterior distribution and  \(\theta \in \{\Omega_m, \Omega_\Lambda\}\) evaluated over 100 test simulations (Table~\ref{tab:priors_training_testing}).  The posterior uncertainty \(\sigma_{\theta}\) is defined as \((P_{84} - P_{16})/2\), where \(P_{16}\) and \(P_{84}\) are the 16th and 84th percentiles of the marginal posterior samples, respectively. Bias is reported in units of \(\sigma_{\theta}\), enabling direct comparison across methods with differing posterior widths. Together, these metrics quantify both the accuracy and precision of the inferred cosmological parameters. In each corner plot, we quote the median as the point estimate together with the 68\% credible interval for each marginalized distribution. In addition, when comparing SBI pipelines we use coverage plots—a standard SBI diagnostic—to test whether the true parameter values fall within the quoted credible intervals at the expected frequencies.

\subsection{Comparison with SALT-SBI and SALT-MCMC}

In this section, we present the median bias and uncertainty obtained from 100 test simulations for each of the three approaches in Table \ref{tab:bias_unc_3}.  (Since SALT-MCMC requires original (uncompressed) $m_B, x_1, c$, we do not quote any encoder-based results until Section \ref{sec:encoder}).  From the tabulated results, the bias ($\sim$ 0.3) and  uncertainty in \(\Omega_\Lambda\) are almost identical for all three approaches. The negative bias observed for the SBI-based methods can be explained by the mismatch between the training and testing priors: the training simulations span \((0, 1)\), while the test simulations are restricted to \((0.4, 1)\) (Table~\ref{tab:priors_training_testing}), so some predicted values are expected to be lower than the smallest value used to generate the test simulations. For consistency, we adopt the same priors for SALT-MCMC as for the SBI methods.

For \(\Omega_m\), SALT-SBI achieves a much smaller bias ($ 0.12$) than FPCA-SBI ($0.30$), while their posterior uncertainties are comparable ($\sim$ 0.20); in this sense, SALT-SBI performs better on in-domain simulations. However, since the light curves are derived from a SALT2-extended SED, strong in-domain performance from SALT2 fitting is expected. FPCA-SBI achieving comparable constraints, despite not sharing this built-in advantage, is a noteworthy  outcome. The positive bias can once again be explained by the priors in training \((0, 1)\) vs test (\(0, 0.6)\) simulations, which causes some of the predicted values to be higher than the largest value adopted for test simulations.  

SALT-MCMC exhibits a  larger bias ($0.56$) than both SBI approaches. This can be attributed to two factors. First, we model the detection probability as a function of redshift only, estimated as the fraction of SALT-fitted supernovae over the total number simulated at a given redshift. We do not explicitly model the underlying $3\sigma$ detection threshold or the quality cuts applied during SALT fitting. Second,  for the suite of 100 test simulations, the Tripp-relation nuisance parameters, $M_B$, \(\alpha\) and \(\beta\), are drawn from Gaussian distributions, and in the SALT-MCMC analysis we adopt the same Gaussian priors. In this sense, the priors are correctly specified at the population level. However, for any individual realisation, the true values, \(M_{B,\mathrm{true}},\, \alpha_{\mathrm{true}},\, \beta_{\mathrm{true}}\), need not coincide with the mean of the prior. To probe the impact of these choices,  we evaluate SALT-MCMC by analysing a single simulation generated with \(\alpha\) and \(\beta\) under this controlled setting, the recovered cosmological parameters show noticeably reduced bias discrepancy---  $0.39$ using SALT-MCMC vs $0.34$ using FPCA-SBI --- highlighting that the MCMC estimates are highly sensitive to the assumed priors. The remaining discrepancy can be attributed to the simplified modeling of the detection probability.

\begin{table*}
\centering
\begin{tabular}{c | c| c| c| c}
\hline
Method  & bias/$\sigma$ ($\Omega_{m}$) & bias/$\sigma$ ($\Omega_{\Lambda}$)  & 1-$\sigma$ ($\Omega_{m}$) & 1-$\sigma$ ( $\Omega_{\Lambda}$) \\
\hline
FPCA-SBI & 0.30 & -0.32 & 0.18 & 0.21\\
\hline
SALT-SBI& 0.12 & -0.30 & 0.20 & 0.21\\
\hline
SALT-MCMC& 0.56 & -0.33 & 0.11 &0.26 \\
\hline
\end{tabular}

\caption[Bias and uncertainty: FPCA-SBI vs. SALT-SBI and SALT-MCMC]{Comparison of the median posterior bias and uncertainty across 100 test simulations for FPCA-SBI, SALT-SBI, and SALT-MCMC.}
\label{tab:bias_unc_3}
\end{table*}

\begin{figure}[h]
    \centering
        \includegraphics[scale=0.6]{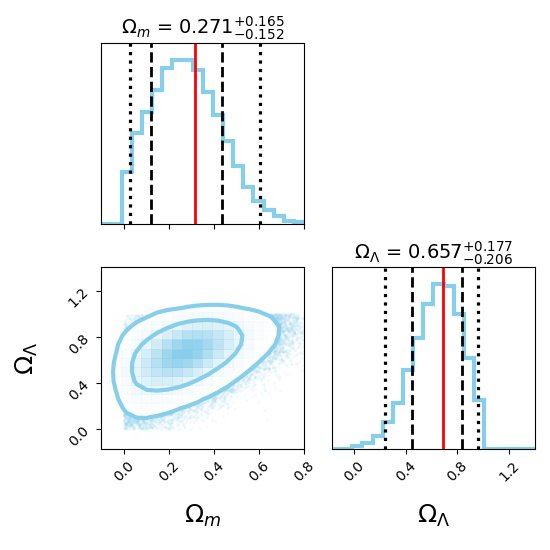}
        \caption[Posterior distribution of  \(\Omega_m\) and \(\Omega_\Lambda\)  for one test simulation generated with Planck cosmological parameters using FPCA-SBI]{Posterior distributions of \(\Omega_m\) and \(\Omega_\Lambda\) from one test simulation generated with input Planck 2018 cosmology (using FPCA-SBI). The contour shows the joint posterior, and the marginalised histograms are shown on the diagonal panels. The black dashed and dotted lines indicate the 1-\(\sigma\)  and 2-\(\sigma\) confidence intervals respectively. The true Planck values (\(\Omega_m = 0.315\) and \(\Omega_\Lambda = 0.685\)) are marked in red and fall within the 1-\(\sigma\) region for both parameters.}
    \label{fig:corner_pc_single}
\end{figure}

In Figure \ref{fig:corner_pc_single}, we plot the joint and marginal posteriors for the inferred cosmological parameters  for  one set of test simulations generated with best-fit  Planck 2018 cosmology ($\Omega_m=0.315$, $\Omega_\Lambda=0.685$) using FPCA-SBI.  The contour (off-diagonal plot) represents the joint posterior, and marginalized histograms for each parameter are shown on the diagonal panels. The true values are shown by red solid lines, and black dashed and dotted lines indicate the 1-$\sigma$ (68\%) and 2-$\sigma$ (95\%) credible regions respectively.  Hence, the true Planck values  fall within the 1-$\sigma$ region for both parameters.

\begin{figure}
    \centering
        \includegraphics[scale=0.6]{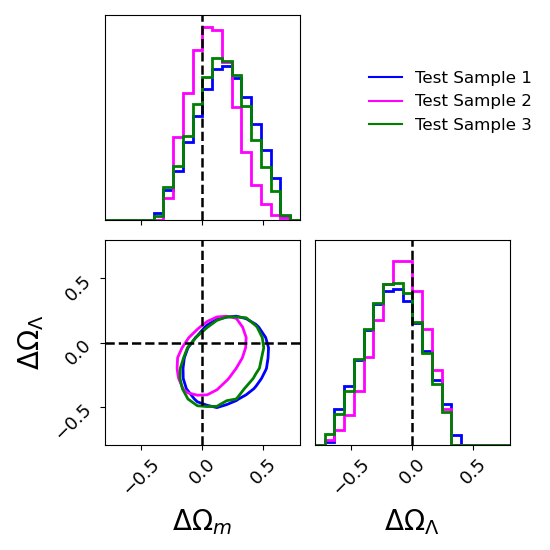}
        \caption[Deviation of posterior distribution of  \(\Omega_m\) and \(\Omega_\Lambda\) for three test simulations  using FPCA-SBI]{Deviations of posterior distributions of \(\Omega_m\) and \(\Omega_\Lambda\) from three test simulations using FPCA-SBI.  The true  values (zero deviations) are shown in black dashed lines and  fall within the 1-\(\sigma\) region for both parameters for all three simulations.}
    \label{fig:corner_pc_multiple}
\end{figure}

In Figure \ref{fig:corner_pc_multiple}, we plot the joint and marginal posterior deviations for three test simulations using FPCA-SBI. The true cosmological parameter values for these simulations are listed in Table~\ref{tab:3_cosmo_true}, spanning flat (simulation 1), positively curved (simulation 2), and negatively curved (simulation 3) universes. For each posterior sample we compute the deviation by subtracting the true value, so that the true cosmology corresponds to zero deviation, shown by the black dashed lines, and  is recovered within the 1-$\sigma$ region for both parameters for all three simulations.

\begin{table}
\centering
\begin{tabular}{c| c| c}
\hline
$ $ & $\Omega_{m}$ & $\Omega_{\Lambda}$ \\
\hline
 Simulation 1& 0.35 & 0.65 \\
 Simulation 2& 0.27 & 0.70 \\
 Simulation 3& 0.34 & 0.69 \\
\hline
\end{tabular}
\caption{True cosmology value of the three simulations shown in Figure \ref{fig:corner_pc_multiple}}
\label{tab:3_cosmo_true}
\end{table}

\begin{figure}[h]
    \centering
        \includegraphics[scale=0.6]{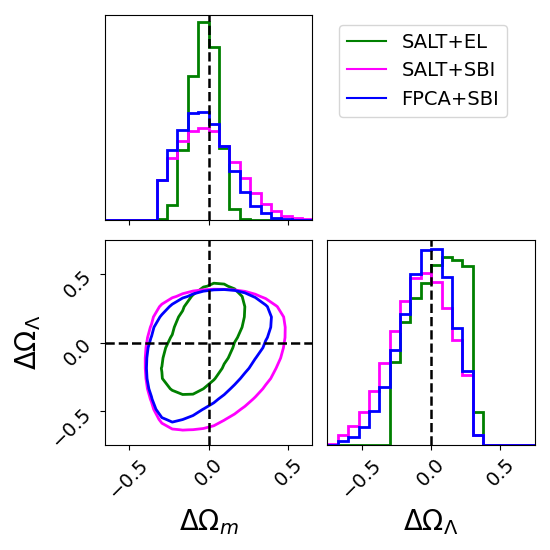}
        \caption[Deviation of posterior distribution of  \(\Omega_m\) and \(\Omega_\Lambda\)  for one test simulation generated with Planck cosmological parameters using three methods]{Deviations of posterior distributions of \(\Omega_m\) and \(\Omega_\Lambda\) from one test simulation   generated with input Planck 2018 cosmology using three methods.  The true  values (zero deviations) are shown in black dashed lines and  fall within the 1-\(\sigma\) region for both parameters for all three simulations.}
    \label{fig:corner_all_multiple}
\end{figure}

In Figure \ref{fig:corner_all_multiple}, we plot the joint and marginal posterior deviations for one simulation  generated with Planck 2018 cosmology ($\Omega_m=0.315$, $\Omega_\Lambda=0.685$) using three methods. SALT-MCMC, SALT-SBI and FPCA-SBI results are shown in green, red and blue respectively. The 1$\sigma$ uncertainty for $\Omega_m$ using SALT-MCMC is smaller than the SBI-based methods, which results in its narrower posterior width. As in Figure \ref{fig:corner_pc_multiple}, deviation is computed by  subtracting the true value from each posterior sample, so that the true cosmology corresponds to zero deviation, shown by the black dashed lines, and  is recovered within the 1-$\sigma$ limit for all three methods.

\subsection{K-correction}
K-corrections \citep{hogg2002k} are used to account for differences in filter bandpasses and for the redshifting of the supernova SED. In our setup, redshift is already incorporated as an explicit input feature to the MDN, and the training and test simulations are generated using the same LSST \(i\) and \(z\) filters. In this configuration, the network can, in principle, learn the redshift and filter dependence directly from the simulations, so explicit K-corrections are not strictly required for internal consistency.

To assess the impact of K-corrections, we apply them  to the fitted peak magnitudes   for a single observed filter;  we  do not perform color matching via spectrum mangling. We deliberately avoid K-correcting the raw light curves, since this would require additional spectral assumptions at each epoch and can remove information that FPCA might otherwise exploit. Instead, after light-curve fitting, k-corrections are computed using Hsiao SED templates \citep{hsiao2007k}  as functions of redshift for the chosen filter and apply these directly to the fitted peak magnitude in that band. 

In Table \ref{tab:k_corr} we summarize the bias and uncertainty obtained when the MDN is trained directly on the raw FPCA summaries versus on FPCA summaries constructed from K-corrected peak magnitudes. The mean biases in \(\Omega_m\) and \(\Omega_\Lambda\) change  slightly when k-corrections are incorporated, from \(0.30\) to \(0.35\) for \(\Omega_m\) and from \(-0.32\) to \(-0.38\) for \(\Omega_\Lambda\), and remain consistent within the quoted uncertainties. The small increase in bias is plausibly explained by our simplified K-correction scheme---single filter and no spectrum mangling. Given that redshift is included explicitly as an MDN feature and that all simulations and test data use the same LSST \(i\) and \(z\) filter set, the near-equivalence of the K-corrected and non-corrected results is expected.

Because the K-corrected and uncorrected results are statistically consistent, and the uncorrected pipeline avoids additional spectral assumptions, we adopt the uncorrected FPCA summaries for all subsequent analyses. In future applications that combine light curves from multiple surveys with different filter systems, a more detailed treatment of k-corrections would be required to control inter-survey systematics.

\begin{table*}
\centering
\begin{tabular}{c | c| c| c| c}
\hline
Method  & bias/$\sigma$ ($\Omega_{m}$) & bias/$\sigma$ ($\Omega_{\Lambda}$)  & 1-$\sigma$ ($\Omega_{m}$) & 1-$\sigma$ ( $\Omega_{\Lambda}$) \\
\hline
Direct FPCA & 0.30 & -0.32 & 0.18 & 0.21\\
\hline
K-corrected FPCA& 0.35 & -0.38 & 0.19 & 0.21\\
\hline
\end{tabular}

\caption[Bias and uncertainty: K-corrected vs direct FPCA outputs]{Comparison of the median posterior bias and uncertainty across 100 test simulations for direct FPCA outputs as MDN inputs versus K-corrected FPCA outputs as MDN inputs.}
\label{tab:k_corr}
\end{table*}

\subsection{Encoder Outputs}

One advantage of using summary statistics with a machine learning model is that the summary statistic vector can be compressed into a lower-dimensional space, reducing training time. More critically, in out-of-domain regimes this compression can mitigate convergence issues  by projecting into a more compact, information-dense representation, the encoder can partially overcome domain mismatch limitations (see Section~\ref{sec:ood}).

In Table~\ref{tab:bias_unc_encoder} we report the median bias and $1\sigma$ uncertainties for 100 test samples, with and without the encoder. We retain the 50-dimensional latent space encoder described in Section~\ref{sec:encoder}. 
Results without the encoder are given in parentheses. The $1\sigma$ uncertainties are comparable between encoded and non-encoded inputs. For FPCA-SBI, the $\Omega_m$ bias reduces by roughly two-thirds, from $0.32$ to $0.10$, while the $\Omega_\Lambda$ bias remains unchanged. For SALT-SBI, the 
$\Omega_m$ bias is unchanged while the $\Omega_\Lambda$ bias reduces from $-0.30$ to $-0.04$. In both cases, the encoder reduces bias in one parameter without degrading the other, demonstrating that the compact latent representation either 
improves or preserves inference performance.

\begin{table*}
\centering
\begin{tabular}{c | c| c| c| c}
\hline
Method  & bias/$\sigma$ ($\Omega_{m}$) & bias/$\sigma$ ($\Omega_{\Lambda}$)  & 1-$\sigma$ ($\Omega_{m}$) & 1-$\sigma$ ( $\Omega_{\Lambda}$) \\
\hline
FPCA-SBI & 0.10 (0.32) & -0.34 (-0.32) & 0.21(0.18) & 0.28 (0.21)\\
\hline
SALT-SBI& 0.12 (0.12) & -0.04(-0.30) & 0.16(0.20) & 0.24(0.21)\\
\hline
\end{tabular}
\caption{Comparison of  posterior bias and uncertainty across 100 test simulations for  encoded vs non-encoded inputs. The number in parenthesis are for non-encoded inputs referenced for comparison. Using an encoder does not change the inference confidence, but improves the bias in one parameter for both methods.}
\label{tab:bias_unc_encoder}
\end{table*}

\subsection{SBI Diagnostic}
In this section, we describe two commonly used SBI diagnostics: the coverage plot and the rank plot. Both measure how well the posteriors are calibrated. For this evaluation, we use a separate test set: we generate 200 additional test simulations with the same distributions as the training set defined in Table \ref{tab:priors_training_testing}, because these metrics are defined under the assumption that training and testing simulations follow the same distributions.

\subsubsection{Coverage Plot}

Coverage plots provide a standard diagnostic for SBI performance, answering the question: how often is the true parameter value recovered within a given credible interval? For $n$ test samples, each with a known truth, we compute the fraction of posterior draws that bracket the truth at each credible level, and plot this 
fraction against the nominal credible level.

\begin{figure*}[htbp]
\centering
\begin{subfigure}[t]{0.48\textwidth}
    \centering
    \includegraphics[width=\textwidth]{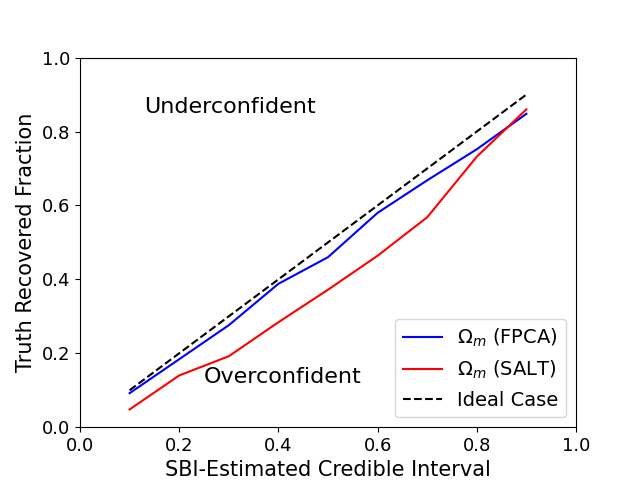}
    \label{fig:coverage_m}
\end{subfigure}
\hfill
\begin{subfigure}[t]{0.48\textwidth}
    \centering
    \includegraphics[width=\textwidth]{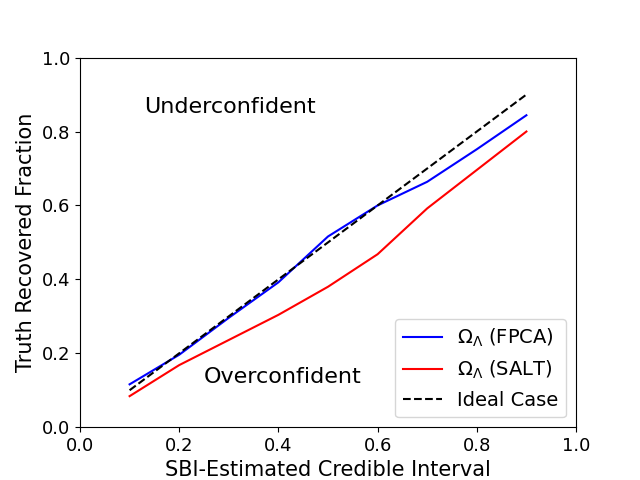}
     \label{fig:coverage_l}
\end{subfigure}
\caption{Coverage plots for $\Omega_m$ (left) and $\Omega_\Lambda$ (right). The black diagonal line represents perfect calibration. FPCA-SBI (blue) and SALT-SBI (red) are shown for comparison. For both parameters, the FPCA-SBI posterior is better calibrated than its SALT counterpart, though the differences are modest. }
\label{fig:coverage_plots}
\end{figure*}

Figure~\ref{fig:coverage_plots} shows the coverage plots for $\Omega_m$ (left) and $\Omega_\Lambda$ (right). The $x$-axis shows the nominal credible level (the central fraction of posterior samples considered), and the $y$-axis shows 
the empirical coverage (the fraction of test cases in which the truth falls within that interval). A perfectly calibrated posterior follows the diagonal one-to-one line (black). Deviations above the diagonal indicate underconfidence (posteriors too wide), while deviations below indicate overconfidence (posteriors too narrow).

FPCA-SBI results are shown in blue and SALT-SBI in red. For both $\Omega_m$ and $\Omega_\Lambda$, the blue curve lies closer to the ideal diagonal than the red curve, indicating that the posteriors obtained from the encoded FPCA-based summary statistics are better calibrated than those from the SALT counterpart, but the differences are not significant.  The FPCA-SBI curve closely traces the diagonal for credible levels below 0.8, with only slight deviations at higher credible levels.

\subsubsection{Rank Distribution}
The rank plot is another diagnostic for assessing how well the posterior is calibrated.  For each of 200 test samples, the true parameters $(\Omega_m, \Omega_\Lambda)$ are drawn from the prior, and the corresponding summary statistics are simulated. The trained posterior is then conditioned on these summary statistics to draw $N$ posterior samples. The rank is defined as the number of those posterior samples whose parameter value falls below the true value. This procedure is repeated for all 200 test samples, and the resulting ranks are plotted as a histogram of 10 bins. Following \citet{talts2018validating}, a perfectly-calibrated posterior should produce ranks that are uniformly distributed. 

\begin{figure*}[htbp]
    \centering
    \includegraphics[scale=0.42]{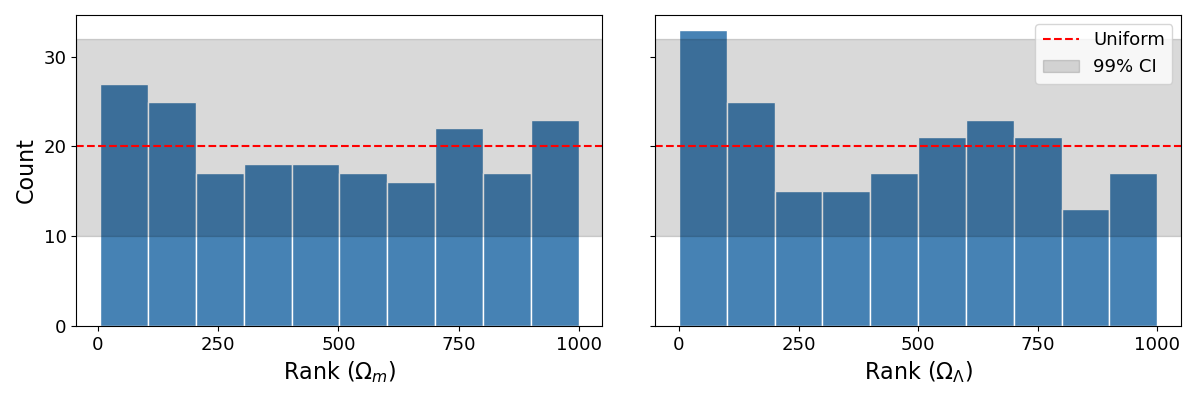}
    \caption{Rank plot showing the distribution of }
    \label{fig:rank}
\end{figure*}

We show the results in Figure~\ref{fig:rank}. The red dashed line indicates the expected count under perfect uniformity ($200~\text{samples} / 10~\text{bins} = 20$ per bin), and the grey shaded region denotes the 99\% confidence interval. All 10 bins for both $\Omega_m$ and $\Omega_\Lambda$ fall within this confidence interval, indicating no statistically significant deviation from uniformity. Some degree of fluctuations around the expected count are attributable to sampling noise, as 200 test samples is a relatively modest number for rank-based calibration tests.

\subsection{Out-of-domain tests}
\label{sec:ood}
One of the major drawbacks of SBI --- and machine learning models more broadly --- is performance degradation when there is a mismatch between training and test distributions. This is particularly relevant here because different photometric surveys exhibit very different observational characteristics. More specifically, the global nuisance parameters $\alpha$ and $\beta$ in the Tripp relation are known to take different values when derived from different datasets \citep{betoule2014improved}.

To assess robustness to such distribution shifts, we perform two out-of-domain tests. In the first test, we shift the mean of the $\alpha$ distribution between training and test sets --- from $\mu_{\alpha} = 0.167$ to $\mu_{\alpha} = 1.5$ --- while keeping the $\beta$ distribution fixed across both sets (first row of Table~\ref{tab:priors_out_of_domain}). In the second test, we fix $\alpha$ to the same distribution ($\mu_{\alpha} = 0.167$) in both sets but shift the mean of the $\beta$ distribution from $\mu_{\beta} = 3.51$ in the training set to $\mu_{\beta} = 2.70$ in the test set (second row of Table~\ref{tab:priors_out_of_domain}). In both tests, all distributions are modeled as truncated Gaussians.

\begin{table*}
\centering
\begin{tabular}{c|c|c}
\hline
Parameter & Prior (Training) & Prior (Test)\\
\hline
$\alpha$ & $\mathcal{TN}(0.167,\,0.012;\,0.131,\,0.203)$ 
& $\mathcal{TN}(1.5,\,0.1;\,1.3,\,1.7)$\\
\hline
$\beta$ & $\mathcal{TN}(3.51,\,0.16;\,3.03,\,3.99)$
&  $\mathcal{TN}(2.70,\,0.10;\,2.50,\,2.90)$ \\
\hline
\end{tabular}
\caption{Prior distributions of $\alpha$ and $\beta$  for out-of-domain generalization tests. The training and test sets are drawn from deliberately mismatched priors to evaluate the robustness of the summary statistics derived from each method. As before $\mathcal{TN}$ denotes
a Truncated Normal distribution.}
\label{tab:priors_out_of_domain}
\end{table*}

\begin{table}
\centering
\begin{tabular}{c | c| c}
\hline
Parameter & bias/$\sigma$ ($\Omega_{m}$) & bias/$\sigma$ ($\Omega_{\Lambda}$)   \\
\hline
$\alpha$ & 2.74 & -3.25 \\
\hline
$\beta$ & -0.57 & 0.01 \\
\hline
Unbiased & 0.10  & -0.34  \\
\hline
\end{tabular}
\caption{Median posterior bias and uncertainty across 100 test simulations for  out-of-domain tests for FPCA-SBI. For reference, the in-domain results are given on the last row.}
\label{tab:bias_unc_out-of_domain_fpca}
\end{table}

\begin{table}
\centering
\begin{tabular}{c | c| c}
\hline
Parameter & bias/$\sigma$ ($\Omega_{m}$) & bias/$\sigma$ ($\Omega_{\Lambda}$)   \\
\hline
$\alpha$ & 5.59  & -4.10  \\
\hline
$\beta$ & -0.55 & -0.05 \\
\hline
Unbiased & 0.12  & -0.04  \\
\hline
\end{tabular}
\caption{Median posterior bias and uncertainty across 100 test simulations for  out-of-domain tests for SALT-SBI. For reference, the in-domain results are given on the last row.}
\label{tab:bias_unc_out-of_domain_salt}
\end{table}

It is worth noting that for the $\beta$-shift test, both FPCA and SALT2 converge with unencoded inputs. However, when the distribution of $\alpha$ is shifted, the posteriors fail to converge for both methods when using unencoded inputs, further motivating the use of the encoder for out-of-domain robustness.

Examining Tables~\ref{tab:bias_unc_out-of_domain_fpca} and 
\ref{tab:bias_unc_out-of_domain_salt}, we find that when $\beta$ is shifted, the out-of-domain bias is comparable between FPCA and SALT2 inputs. However, when $\alpha$ is shifted, the bias in $\Omega_m$ is 2.74 for FPCA versus 5.59 for SALT2 --- more than double --- despite both methods having nearly identical in-domain biases of 0.10 and 0.12, respectively. Similarly, for $\Omega_\Lambda$, 
the out-of-domain bias is ${\sim}{-3}$ for FPCA and ${\sim}{-4}$ for SALT2, even though the in-domain bias was smaller for SALT2.

These results suggest that FPCA-based representations are inherently more robust to nuisance parameter shifts. Unlike SALT2, which applies fixed stretch and color corrections, FPCA relaxes rigid template-fitting assumptions and captures a more diverse range of light curve behaviour. This flexibility appears to confer stronger domain adaptation, suggesting that FPCA-SBI may be better positioned to remain reliable under the varying survey conditions expected for LSST and Roman --- and potentially to probe physics beyond the standard stretch-and-color correction framework.

\begin{figure}[ht!]
    \centering
    \includegraphics[scale=0.62]{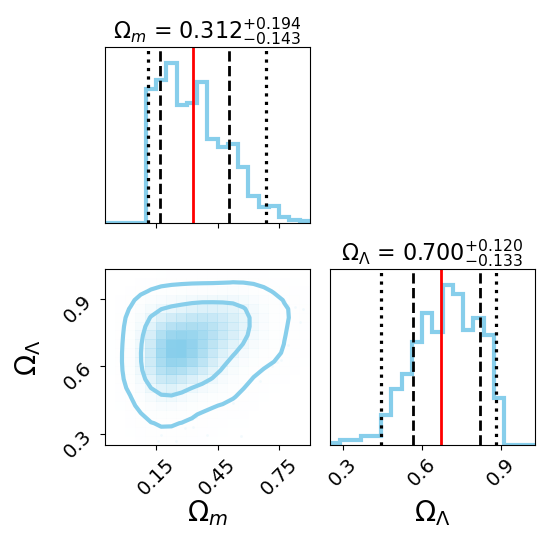}
    \caption{Corner plot showing the joint and marginal distributions of the cosmological parameters for the 204 spectroscopically confirmed DES SNe Ia. The results are consistent with those obtained by the DES team for a non-flat $\Lambda$CDM model.}
    \label{fig:des_corner_plot}
\end{figure}

\subsection{Application to DES supernovae}
In this section, we apply the trained model to the spectroscopically confirmed SNe Ia from the DES Y5 dataset to infer cosmological parameters and compare the results with \citet{abbott2019first} for a non-flat $\Lambda$CDM model. This serves as the first application of the proposed method to real observational data.

Since the model is trained on FPCA parameters derived from simulated LSST light curves in the $i$ and $z$ bands, we fit only $i$ and $z$ band light curves for the DES sample. We apply selection cuts consistent with those adopted for the LSST light curves, retaining 204 out of 207 spectroscopically confirmed SNe Ia. Milky Way foreground extinction corrections are applied following the empirical dust extinction law of \citet{fitzpatrick1999correcting}, with $R_V = 3.1$ and $E(B-V)$  values taken from the DES dataset. To match the redshift distribution of the 204 DES SNe Ia, we select 204 out of 800 SNe from the training simulations based on the closest redshift match. We note that the DES spectroscopic sample is predominantly low-$z$, and the redshift distributions are therefore not closely matched. The encoder is retrained on this redshift-matched training subset of 204 LSST SNe to learn the latent representation. The encoder's latent-space dimensionality is kept unchanged at 50, and the resulting latent variables are used to retrain the Mixture Density Network. This retraining is done only to match the number of SNe between the training and test sets — the model (encoder + Mixture Density Network) is still trained on the simulated LSST SNe and applied to the DES spectroscopic sample.

The corner plot in Figure~\ref{fig:des_corner_plot} shows the joint and marginal posterior distributions of the cosmological parameters. The posterior median yields $\Omega_m = 0.312$ and $\Omega_\Lambda = 0.700$. For reference, the likelihood-based MCMC approach of \citet{abbott2019first} reported median values of $\Omega_m = 0.332 \pm 0.122$ and $\Omega_\Lambda = 0.671 \pm 0.163$ for a non-flat $\Lambda$CDM model, which corresponds  to biases of $-0.12\sigma$ and $0.23\sigma$ for $\Omega_m$ and $\Omega_\Lambda$, respectively between our method and their method. This demonstrates that the model performs satisfactorily on real observational data, with bias values within an acceptable range. However, it is to be noted that with only 204 SNe per simulation, MDN training is not very stable and the results fluctuate with model initialization. For more robust inference, it is imperative to train the network on a sample comparable in size to the full photometrically classified DES dataset, with matched redshift distribution and survey conditions including cadence and filter coverage.

\subsection{Host Galaxy}

As discussed in the Introduction, SN peak luminosity and standardization parameters ($\alpha$, $\beta$) depend on host galaxy properties. While the SALT2 Tripp relation accounts only for stretch and color, it does not intrinsically capture host-galaxy dependence; correcting for known correlations between SN Ia properties and their host environments therefore requires augmenting the standardization with additional, explicitly fitted host-dependent terms. FPCA's flexible, data-driven decomposition, by contrast, may capture subtle host-galaxy-driven variations encoded directly in the light curve morphology, without recourse to such added terms. If so, this would represent a distinct advantage as separate modeling of host galaxy properties would no longer be required. For this proof-of-concept analysis, we extend our FPCA framework from 2 to 4 parameters to test whether FPCA can capture such variations intrinsically, and whether explicitly incorporating host galaxy properties into the trained models reduces systematic bias in cosmological parameter inference.

We perform a test in which host galaxy properties are introduced through their influence on the dust extinction law. For each simulation, the 800 SNe~Ia are divided into two equal subsamples of 400, partitioned by redshift to reflect the observed evolution of host demographics: high-$z$ hosts are predominantly early-type and less massive, whereas low-$z$ hosts are typically more massive. Host stellar mass, $\log(M_*/M_\odot)$, is drawn from a truncated normal distribution, $\mathcal{TN}(\mu = 9,\ \sigma = 1,\ a = 7,\ b = 12)$ for the high-$z$ subsample and $\mathcal{TN}(\mu = 10.5,\ \sigma = 0.8,\ a = 7,\ b = 12)$ for the low-$z$ subsample, where $a$ and $b$ denote the truncation bounds; both distributions are motivated by \citet{sullivan2010dependence}. After stretch and color standardization, SNe~Ia in more massive hosts are observed to be systematically brighter than those in less massive hosts \citep{sullivan2010dependence, kelly2010hubble}. The physical origin of this ``mass step'' remains debated, with proposed explanations spanning differences in progenitor age and metallicity as well as host-dependent dust properties. \citet{brout2021s} attribute the effect primarily to dust, finding that the extinction law differs between high- and low-mass hosts at the $2.9\sigma$ level in $R_V$; we adopt this interpretation to inject a controlled systematic, assigning $R_V \sim \mathcal{U}(1.4,\ 1.6)$ to SNe in massive hosts ($\log(M_*/M_\odot) > 10$) and $R_V \sim \mathcal{U}(2.65,\ 2.85)$ to those in less massive hosts, while holding the color excess fixed at $E(B-V) = 0.15$ in order to isolate the dependence on $R_V$.

\begin{table}
\centering
\begin{tabular}{c|c}
\hline
Parameter & Priors \\
\hline
$\Omega_m$ & $\mathcal{TN}(0.3,\,0.1;\,0.0,\,0.6)$\\
\hline
$\Omega_\Lambda$ & $\mathcal{TN}(0.7,\,0.1;\,0.4,\,1.0)$ \\
\hline
$\alpha$ & $\mathcal{TN}(0.167 0.012;\,0.143,\,0.191)$ \\
\hline
$\beta$ & $\mathcal{TN}(3.51, 0.16;\,3.19,\,3.83)$ \\
\hline
${M_B}$ 
& $\mathcal{TN}(-19.3, 0.1;\,-19.5,\,-19.1)$  \\
\hline
$x_1$ & $\mathcal{TN}(0,\,1;\,-3,\,3)$\\
\hline
$c$ & $\mathcal{TN}(0,\,0.1;\,-0.3,\,0.3)$\\
\hline
\end{tabular}
\caption{Prior distributions of cosmological, nuisance and light curve  parameters for the host mass-dependent dust simulations. Same priors are used for training and test simulations. Here  $\mathcal{TN}$ denotes Truncated Normal distributions, respectively.}
\label{tab:priors_mass_step}
\end{table}

We first perform two statistical tests to assess whether any individual FPCA coefficient differs significantly between high- and low-mass host galaxy groups. The sample is divided into two groups based on host stellar mass: $\log(M_*/M_\odot) > 10$ and $\log(M_*/M_\odot) \leq 10$, consistent with the mass threshold used to assign the host-dependent dust ($R_V$) properties in our simulations. We then compute the Mann-Whitney U and KS test p-values for each FPPC coefficient between the two groups. The Mann-Whitney U test measures differences in the central tendency of the two distributions, while the KS test is a more sensitive metric that compares the full shape of the distributions. A p-value below 0.05 indicates a statistically significant difference between the two groups. Because each test simulation contains 800 SNe~Ia, these metrics are computed over the 800 SNe belonging to a given simulation at once, and we report in Table~\ref{tab:mw_ks_pvalues} the median of the resulting $p$-values across the 100 test simulations. For the first two FPCA scores in both bands, both the Mann-Whitney U and KS $p$-values fall below 0.05, indicating that the host-galaxy signal injected via the mass-dependent dust extinction is statistically imprinted on these leading FPCA features. The higher-order coefficients show no such dependence, suggesting the imprint is confined to the dominant light-curve modes. In appendix \ref{app:hist}, we plot the histogram distributions of the 8 FPCA coefficients for the two mass groups for one randomly selected simulation.

\begin{table}[h!]
\centering

\begin{tabular}{c|c|c}
\hline
FPCA & $p_{\rm MW}$ & $p_{\rm KS}$ \\
\hline
$i\_a_1$ & $0.013$ & $0.018$ \\
$i\_a_2$ & $0.008$ & $0.013$ \\
$i\_a_3$ & $0.438$ & $0.402$ \\
$i\_a_4$ & $0.414$ & $0.420$ \\
$z\_a_1$ & $0.015$ & $0.017$ \\
$z\_a_2$ & $0.010$ & $0.010$ \\
$z\_a_3$ & $0.442$ & $0.584$ \\
$z\_a_4$ & $0.414$ & $0.440$ \\
\hline
\end{tabular}
\caption{Mann-Whitney U and KS test p-values for FPCA scores between intrinsically brighter SNe Ia hosted in high-mass ($\log(M_*/M_\odot) > 10$) and dimmer SNe Ia hosted in 
low-mass ($\log(M_*/M_\odot) \leq 10$) galaxies..}
\label{tab:mw_ks_pvalues}
\end{table}

We next compare the posterior bias and uncertainties in cosmological parameters between simulations with and without host stellar mass included as an explicit feature during MDN training. The median bias and uncertainty for 100 test simulations are tabulated in Table \ref{tab:bias_with_without_mass}. From the tabulated results, the 1-$\sigma$ uncertainties of both parameters remain unchanged when the host galaxy mass is supplied to the model. Turning to  bias, incorporating the host mass lowers the  $\Omega_m$ bias from $0.18$ to $0.13$ and increases the $\Omega_\Lambda$ bias from $0.09$ to $0.14$. These shifts are comparable in magnitude and offset one another, and in neither case is the change statistically significant, so we find no net improvement from including host mass. These results suggest that the 4-component FPCA representation intrinsically captures the host-galaxy dependency for this simulation setup, such that explicit separate modeling of the host stellar mass may not be required for unbiased cosmological parameter inference.

\begin{figure*}[htbp]
\centering
\begin{subfigure}[t]{0.48\textwidth}
    \centering
    \includegraphics[width=\textwidth]{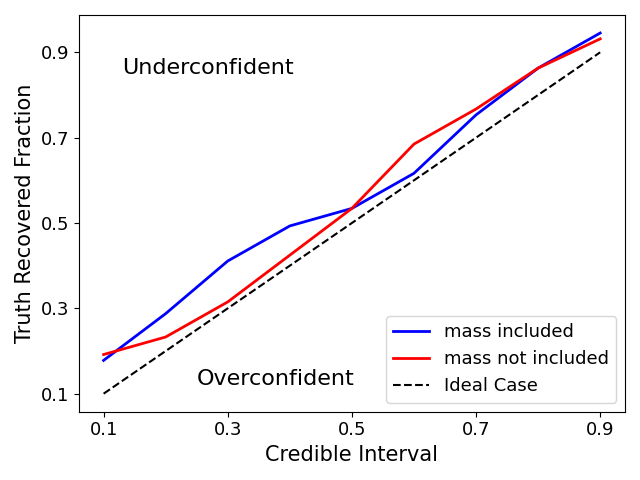}
    \caption{$\Omega_m$}
\end{subfigure}
\hfill
\begin{subfigure}[t]{0.48\textwidth}
    \centering
    \includegraphics[width=\textwidth]{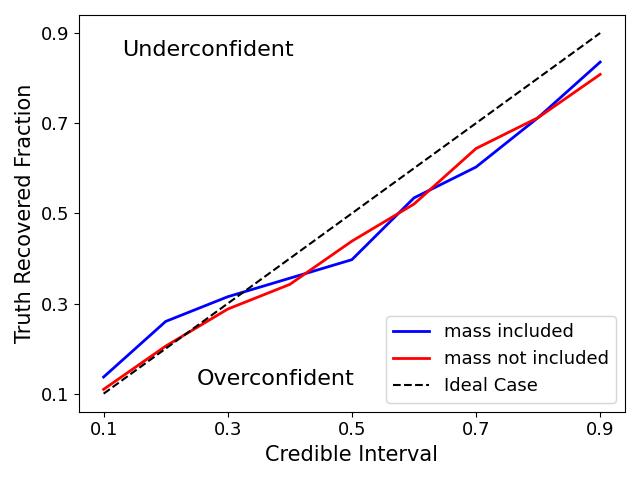}
     \caption{$\Omega_\Lambda$}
\end{subfigure}
\caption{Coverage plots for $\Omega_m$ (left) and $\Omega_\Lambda$ (right). The blue curve corresponds to the case where the host mass is explicitly included during model training, while the red curve corresponds to the 
case where it is not included.}
\label{fig:coverage_plots_mass_step}
\end{figure*}

We next examine the coverage plots in Figure \ref{fig:coverage_plots_mass_step} as an additional visual metric, alongside the bias and uncertainty. The left panel shows the coverage for the matter density $\Omega_m$, and the right panel shows the coverage for the dark-energy density $\Omega_\Lambda$. The black diagonal line indicates the ideal case. The red curve corresponds to the model in which the host mass is not included as an explicit parameter during MDN training, whereas the blue curve corresponds to the model in which it is included.

For $\Omega_m$, the red curve lies closer to the ideal diagonal than the blue curve for credible intervals below $0.5$, after which the ordering reverses. For
$\Omega_\Lambda$, the two curves cross one another several times, indicating that neither shows consistently better coverage across the credible-interval range. Overall, we find no systematic improvement in posterior calibration from including the host mass as an explicit feature. This further strengthens the conclusion from Table~\ref{tab:bias_with_without_mass}, where including the host mass likewise leads to no significant improvement in bias or uncertainty.

\begin{table*}
\centering
\begin{tabular}{c | c| c| c| c}
\hline
Method  & bias/$\sigma$ ($\Omega_{m}$) & bias/$\sigma$ ($\Omega_{\Lambda}$)  & 1-$\sigma$ ($\Omega_{m}$) & 1-$\sigma$ ( $\Omega_{\Lambda}$) \\
\hline
Mass included & -0.13 & -0.14 & 0.08 & 0.10\\
\hline
Mass not included & -0.18 & 0.09 & 0.08 & 0.09\\
\hline
\end{tabular}
\caption{Comparison of the median posterior bias and uncertainty across 100 test simulations when $M_B$ is varied based on host mass.}
\label{tab:bias_with_without_mass}
\end{table*}

\section {Conclusions and Future Work}
\label{sec:conclusions}

We present the first application of Functional Principal Component Analysis (FPCA) in combination with Simulation-Based Inference (SBI) for Type~Ia supernova cosmology. Although the FPCA model was originally developed to characterize the intrinsic spectrophotometric properties of SNe Ia and already have been successful for photometric classification, we extend its utility s here used as a flexible, data-driven summary of light-curve information for cosmological parameter inference. Light curves simulated in the LSST broadband $i$ and $z$ filters, with cadence and noise properties representative of the Vera Rubin Observatory, are fitted using the FPCA fitter. The resulting model parameters are compressed via an autoencoder to $1/100$ of the original dimensionality, and the latent variables are passed to a Mixture Density Network (MDN) to learn the posterior mapping between summary statistics and cosmological parameters. This approach relaxes the simplifying assumptions required to evaluate the likelihood in closed form, offering a principled advantage over traditional MCMC-based methods.

Prior SBI work in supernova cosmology has relied on the Tripp relation and SALT light curve fitting parameters as summary statistics and  explicit-likelihood based approaches. The FPCA-SBI framework introduced here achieves constraints comparable to both approaches, with a median bias of  $\sim$$0.3\,\sigma$ and posterior uncertainties of $\sim$$0.2$ in both $\Omega_m$ and $\Omega_\Lambda$. Crucially, because FPCA does not rely on rigid template-based assumptions, it generalizes more robustly to out-of-domain simulations---that is, to light curves generated under different values of global nuisance parameters---a property that is essential for application across surveys with varying cadence, filter response, and noise characteristics. Furthermore, fitting with a two-component FPCA model reduces computational cost to approximately one-third of that required by SALT-based approaches, an important practical advantage at survey scale.

To validate on observed data, the model trained on LSST-like simulations is applied to the spectroscopically confirmed DES Year~5 light curves. The inferred constraints on $\Omega_m$ and $\Omega_\Lambda$ differ from those of the DES collaboration by $0.12\,\sigma$ and $0.23\,\sigma$, respectively, for a non-flat $\Lambda$CDM cosmology.  Finally, we investigate whether a 4-parameter FPCA model can internally capture host effects or whether explicit inclusion of host information during training improves performance. Injecting host-dependent systematics through a mass-dependent dust extinction law, we find that the two leading FPCA scores in both bands carry a significant imprint of the host effect ($p \lesssim 0.02$). Moreover, the 4-parameter FPCA model yields comparable bias and uncertainty whether or not host mass is explicitly included, indicating no measurable benefit from explicit host-mass conditioning within our framework.

This analysis presents a promising avenue for supernova cosmology with upcoming large-scale surveys such as the Nancy Grace Roman Space Telescope and the Vera C. Rubin Observatory/LSST, offering a unified framework that proceeds directly from photometric light curves to cosmological parameter constraints. The pipeline is computationally efficient and, by relying on a single internally consistent light curve model throughout, has the potential to reduce systematic biases and uncertainties that can arise when combining disparate modeling assumptions across pipeline stages. The fully data-driven nature of this framework is particularly advantageous in the context of upcoming surveys, where many sources of systematic error will be difficult  to characterize in advance. Ultimately, this framework will offer new insights into the properties of dark energy and the universe’s expansion history by using a statistically rich sample of thousands of supernovae observed by these surveys.

This proof-of-concept naturally points toward several directions for future development. We plan to construct more realistic survey simulations that include host-galaxy dust, cadence variations, and a survey-matched number of observed light curves, and to replace the simple retention of all $3\sigma$ detections with probabilistic selection via Bernoulli trials that emulate survey detection and quality cuts. We will also extend the current LSST $i$- and $z$-band simulations to all available photometric bands for a given survey. On the analysis side, we intend to move from a clean, spectroscopically confirmed SN~Ia sample to a more realistic pipeline that first performs photometric classification and then propagates classification uncertainties into the cosmological inference. In addition, we will explore other cosmological models  beyond the  $\Lambda$CDM case to include  $w$CDM, and time-varying $w$CDM, in order to systematically quantify whether increased model flexibility  yields tighter or more robust constraints on the nature of dark energy.

\section* {Data Availability}
The analysis pipeline and supporting code developed for this work will be made publicly available on GitHub upon acceptance of the manuscript.

\begin{acknowledgments}
\software{astropy \citep{robitaille2013astropy},  
sncosmo \citep{barbary2016sncosmo},
sbi \citep{tejero2020sbi} }

\end{acknowledgments}

\begin{contribution}
Moonzarin Reza contributed to conceptualization, methodology development, interpretation of results, and manuscript preparation. Lifan Wang provided overall supervision, offering technical and academic guidance throughout the project.
\end{contribution}

\appendix

\section{Computational Feasibility of SBI Methods}
\label{app:time_sbi_meth}
\begin{table*}
\centering
\begin{tabular}{c | c |c }
\hline
Method  & Time (Architecture: 1)  & Time (Architecture: 2)\\
\hline
SNPE & 11 s (MDN) & 19 s (MAF) \\
\hline
SNRE& 3.5 hr (MLP) & 5.3 hr (RESNET) \\
\hline
SNLE& 278 hr (MAF) & 458 hr (MADE)  \\
\hline
\end{tabular}
\caption{Comparison of time for different SBI methods and network architectures}
\label{tab:time_meth_archi}
\end{table*}

For this analysis, we adopt Sequential Neural Posterior Estimation (SNPE) with a Mixture Density Network (MDN) as the density estimator. Given the dimensionality of this problem, SNPE is the most  computationally convenient choice within practical time constraints. Table~\ref{tab:time_meth_archi} compares Table~\ref{tab:time_meth_archi} compares the end-to-end computational cost---combining training and inference time---across SBI methods and network architectures, where different architectures are paired with each method based on availability---not all architectures are compatible with all methods.

Sequential Neural Posterior Estimation (SNPE) with both Mixture Density Network (MDN) and Masked Autoregressive Flow (MAF) completes training and inference in seconds, making it several orders of magnitude faster than the alternatives. Sequential Neural Ratio Estimation (SNRE) with both Multi-Layer Perception (MLP) and Residual Network (RESNET) requires a few hours hours, while Sequential Neural Likelihood Estimation (SNLE) with both Masked Autoregressive Flow (MAF) and Masked Autoencoder for Distribution Estimation (MADE) requires hundreds of hours.  (For SNLE, we estimated its total runtime by measuring the time to collect 100 posterior samples and linearly scaling to the $\sim$10,000 samples required for inference). 

The speed hierarchy across methods reflects their fundamental algorithmic differences. SNPE learns the posterior directly as an explicit density, requiring no MCMC sampling at inference time, making it naturally fast and well-suited to high-dimensional problems. SNRE, by contrast, trains a binary classifier using cross-entropy loss to estimate likelihood-to-evidence ratios, which is more expensive but still tractable. SNLE is the most computationally demanding: it trains a normalising flow over the full data-vector space, requiring the computation of Jacobian determinants at every training step---an operation whose cost scales poorly with dimensionality, rendering it impractical here.

\section{Host-Mass Dependence of FPCA Scores}
\label{app:hist}

\begin{figure*}
    \centering
    \includegraphics[width=\linewidth]{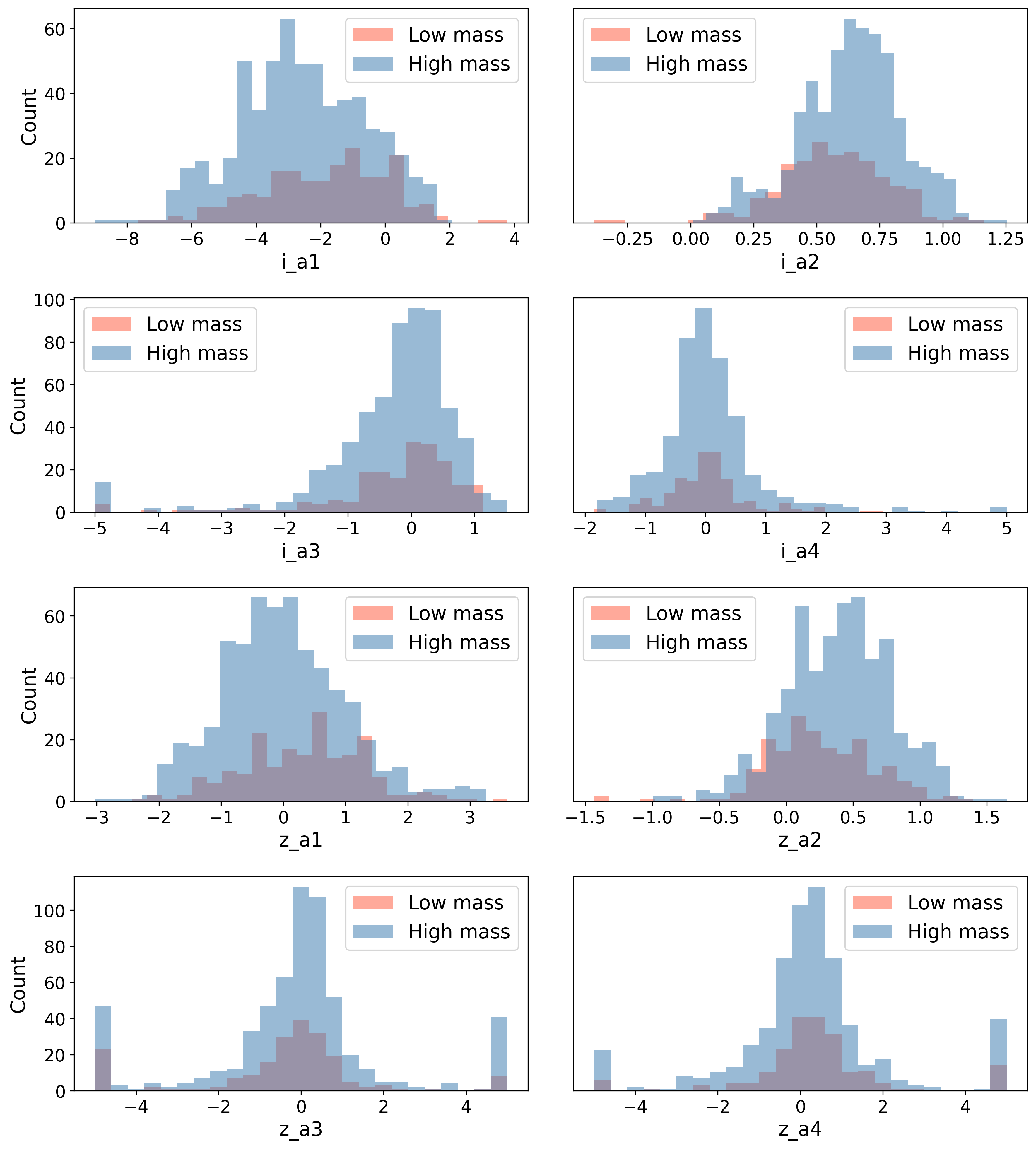}
    \caption{Histogram distributions of the 8 FPCA coefficients for the two mass groups. Low-mass groups are shown in orange and high-mass groups are shown in blue. The shapes of the distributions for the two groups vary widely for the first two coefficients in both bands (first and third rows), while for the higher-order coefficients, the shapes resemble one another.}
    \label{fig:hist_coeff}
\end{figure*}

In this section, we look at the distributions of the 8 FPCA coefficients for the two mass groups, and examine whether they differ. In Figure \ref{fig:hist_coeff}, we plot the histograms of the 8 FPCA coefficients of 800 SNe Ia for one randomly selected simulation, where orange represents the low-mass groups and blue represents the high-mass groups. Since high-mass host galaxies are modeled with lower dust extinction coefficients, their SNe suffer less dimming and more readily pass the signal-to-noise detection cuts, and hence dominate the FPCA-fitted population. As shown in Table \ref{tab:mw_ks_pvalues}, the p-values are significant for i\_${a_1}$, i\_${a_2}$, z\_${a_1}$, z\_${a_2}$. From the figure, we see that the shapes for the two mass groups differ significantly for these four coefficients (first and third rows), while for the other four coefficients (second and fourth rows), the basic shapes between the two groups are comparable, concordant with that result.

\bibliography{sample701}{}
\bibliographystyle{aasjournalv7}

\end{document}